\documentclass[12pt]{article}
\usepackage{graphicx,amsmath}
\usepackage{bm}
\usepackage{color}
\newcounter{mysubsubsection}

\definecolor{db}{rgb}{0.180,0.543,0.340}

\newcommand{\spc}{{\ }}
\newcommand{\pr}[1]{{\sc{\lowercase{#1}}}}

\newcommand{\gras}[1]{\boldsymbol{#1}}

\newcommand{\codeversion}{2.00d}

\newcounter{leteq}

\newenvironment{eqnalpha}{\setcounter{leteq}{1}

\begin{eqnarray}}{\end{eqnarray}%
}

\newenvironment{eqnalphalabel}[1]{\setcounter{leteq}{1}
\raisebox{0cm}[0cm][0cm]{\begin{minipage}{1cm}%
\begin{eqnarray}\label{#1}&&\nonumber\end{eqnarray}\end{minipage}}

\begin{eqnarray}}{\end{eqnarray}%
}

\newcommand{\bnl}{\begin{eqnalpha}}
\newcommand{\enl}{\end{eqnalpha}}
\newcommand{\bnll}[1]{\begin{eqnalphalabel}{#1}}
\newcommand{\enll}{\end{eqnalphalabel}}

\newcommand{\keyw}{{\bf Keyword:}}
\newcommand{\key}[1]{\vspace{1ex}\noindent\keyw{\spc}{\tk{#1}}
                         \newline\phantom{\keyw{\spc}{\tk{#1}}}{\spc}}

\newcommand{\be}{\begin{equation}}
\newcommand{\ee}{\end{equation}}
\newcommand{\ba}{\begin{array}}
\newcommand{\ea}{\end{array}}
\newcommand{\bn}{\begin{eqnarray}}
\newcommand{\en}{\end{eqnarray}}
\newcommand{\bc}{\begin{center}}
\newcommand{\ec}{\end{center}}
\newcommand{\bi}{\begin{itemize}}
\newcommand{\ei}{\end{itemize}}

\newcommand{\tv}[1]{{\tt{#1}}\index{#1}}
\newcommand{\tk}[1]{{\tt{#1}}\index{#1}}

\renewcommand{\tv}[1]{\textcolor{red}    {{\tt{#1}}}}
\renewcommand{\tk}[1]{\textcolor{blue}   {{\tt{#1}}}}

\renewcommand{\tv}[1]{{\tt{#1}}}
\renewcommand{\tk}[1]{{\tt{#1}}}

\begin{document}

\vspace{0.5cm}
\begin{center}
        {\bf\Large
                     Axially deformed solution of the Skyrme-Hartree-Fock-Bogolyubov equations using the transformed harmonic oscillator basis \\[1ex]
                     (II) \pr{hfbtho} v{\codeversion}: a new version of the program.
        }

\vspace{5mm}
        {\large
                       M.V. Stoitsov,$^{a,b}$
                       N. Schunck,$^{c}$\footnote
                       {E-mail: schunck1@llnl.gov}
                       M. Kortelainen,$^{a,b,d}$
                       N. Michel,$^{a}$
                       H. Nam,$^{b}$
                       E. Olsen,$^{a}$
                       J. Sarich,$^{e}$
                       S. Wild,$^{e}$
        }

\vspace{3mm}
        {\it
          $^a$Department of Physics and Astronomy, University of Tennessee,
              Knoxville, TN 37996, USA                                      \\
          $^b$Oak Ridge National Laboratory, P.O. Box 2008,
              Oak Ridge, TN 37831, USA                                      \\
          $^c$Physics Division, Lawrence Livermore National Laboratory
              Livermore, CA 94551, USA                                      \\
          $^d$Department of Physics, P.O. Box 35 (YFL),
              FI-40014 University of Jyv\"askyl\"a, Finland                 \\
          $^e$Mathematics and Computer Science Division, Argonne National Laboratory,\\
              Argonne, IL 60439, USA
        }
\end{center}

\vspace{5mm}
\hrule

\vspace{2mm}
\noindent{\bf Abstract}

We describe the new version {\codeversion} of the code \pr{hfbtho} that solves
the nuclear Skyrme Hartree-Fock (HF) or Skyrme Hartree-Fock-Bogolyubov (HFB)
problem by using the cylindrical transformed deformed harmonic oscillator basis.
In the new version, we have implemented the following features:
(i)   the modified Broyden method for non-linear problems,
(ii)  optional breaking of reflection symmetry,
(iii) calculation of axial multipole moments,
(iv) finite temperature formalism for the HFB method,
(v)  linear constraint method based on the approximation of the Random Phase
Approximation (RPA) matrix for multi-constraint calculations,
(vi)   blocking of quasi-particles in the Equal Filling Approximation (EFA),
(vii)  framework for generalized energy density with arbitrary density-dependences, and
(viii) shared memory parallelism via OpenMP pragmas.
\vspace{2mm}
\hrule

\vspace{2mm}
\noindent
PACS numbers: 07.05.T, 21.60.-n, 21.60.Jz

\vspace{5mm}
{\bf\large NEW VERSION PROGRAM SUMMARY}

\bigskip\noindent{\it Title of the program:} \pr{hfbtho}
                  v{\codeversion}

\bigskip\noindent{\it Catalogue number:}
                   ....

\bigskip\noindent{\it Program obtainable from:}
                      CPC Program Library, \
                      Queen's University of Belfast, N. Ireland
                      (see application form in this issue)

\bigskip\noindent{\it Reference in CPC for earlier version of program:}
                      M.V. Stoitsov, J. Dobaczewski, W. Nazarewicz, P. Ring,
                      Comput.\ Phys.\ Commun.\ {\bf 167} (2005) 43-63.

\bigskip\noindent{\it Catalogue number of previous version:}
                  ADFL\_v2\_1

\bigskip\noindent{\it Licensing provisions:} GPL v3

\bigskip\noindent{\it Does the new version supersede the previous one:} Yes

\bigskip\noindent{\it Computers on which the program has been tested:}
                      Intel Pentium-III, Intel Xeon, AMD-Athlon, AMD-Opteron,
Cray XT5, Cray XE6

\bigskip\noindent{\it Operating systems:} UNIX, LINUX, Windows$^{\text{xp}}$

\bigskip\noindent{\it Programming language used:} FORTRAN-95

\bigskip\noindent{\it Memory required to execute with typical data:} 200 Mwords

\bigskip\noindent{\it No. of bits in a word:} 8

\bigskip\noindent{\it Has the code been vectorised?:} Yes

\bigskip\noindent{\it Has the code been parallelized?:} Yes

\bigskip\noindent{\it No.{\spc}of lines in distributed program:}
                      11 387

\bigskip\noindent{\it Keywords:}
                      Hartree-Fock; Hartree-Fock-Bogolyubov; Nuclear many-body problem;
                      Skyrme interaction; Self-consistent mean field; Density functional theory;
                      Generalized energy density functional; Nuclear matter;
                      Quadrupole deformation; Octupole deformation;
                      Constrained calculations; Potential energy surface; Pairing; Particle number projection;
                      Nuclear radii; Quasiparticle spectra; Harmonic oscillator; Coulomb field;
                      Transformed harmonic oscillator; Finite temperature; Shared memory parallelism.

\bigskip\noindent{\it Nature of physical problem}

\noindent
The solution of self-consistent mean-field equations for weakly-bound paired
nuclei requires a correct description of the asymptotic properties of nuclear
quasiparticle wave functions. In the present implementation, this is achieved
by using the single-particle wave functions of the transformed harmonic oscillator,
which allows for an accurate description of deformation effects and pairing
correlations in nuclei arbitrarily close to the particle drip lines.

\bigskip\noindent{\it Method of solution}

\noindent
The program uses the axial Transformed Harmonic Oscillator (THO) single-particle
basis to expand quasiparticle wave functions. It iteratively diagonalizes the
Hartree-Fock-Bogolyubov Hamiltonian based on generalized Skyrme-like energy
densities and zero-range pairing interactions until a self-consistent solution is
found. A previous version of the program was presented in: M.V. Stoitsov,
J. Dobaczewski, W. Nazarewicz, P. Ring, Comput.\ Phys.\ Commun.\ {\bf 167} (2005)
43-63.

\bigskip\noindent{\it Summary of revisions}

\noindent
\begin{enumerate}
\setlength{\itemsep}{-1ex}
\item The modified Broyden method has been implemented,
\item Optional breaking of reflection symmetry has been implemented,
\item The calculation of all axial multipole moments up to $\lambda = 8$ has been
implemented,
\item The finite temperature formalism for the HFB method has been implemented,
\item The linear constraint method based on the approximation of the Random Phase
Approximation (RPA) matrix for multi-constraint calculations has been implemented,
\item The blocking of quasi-particles in the Equal Filling Approximation (EFA) has
been implemented,
\item The framework for generalized energy density functionals with arbitrary
density-dependence has been implemented,
\item Shared memory parallelism via OpenMP pragmas has been implemented.
\end{enumerate}

\bigskip\noindent{\it Restrictions on the complexity of the problem}

\noindent
Axial- and time-reversal symmetries are assumed.

\bigskip\noindent{\it Typical running time}

\noindent
Highly variable, as it depends on the nucleus, size of the basis, requested 
accuracy, requested configuration, compiler and libraries, and hardware 
architecture. An order of magnitude would be a few seconds for ground-state 
configurations in small bases $N_{\text{max}}\approx 8-12$, to a few minutes 
in very deformed configuration of a heavy nucleus with a large basis 
$N_{\text{max}}> 20$.

\bigskip\noindent{\it Unusual features of the program}

\noindent
The user must have access to (i) the LAPACK subroutines \pr{DSYEVD}, \pr{DSYTRF}
and \pr{DSYTRI}, and their dependencies, which compute eigenvalues and
eigenfunctions of real symmetric matrices, (ii) the LAPACK subroutines \pr{DGETRI}
and \pr{DGETRF}, which invert arbitrary real matrices, and (iii) the BLAS routines
\pr{DCOPY}, \pr{DSCAL}, \pr{DGEMM} and \pr{DGEMV} for double-precision linear
algebra (or provide another set of subroutines that can perform such tasks). The
BLAS and LAPACK subroutines can be obtained from the Netlib Repository at the
University of Tennessee, Knoxville: \verb+http://netlib2.cs.utk.edu/+.

\bigskip

{\bf\large LONG WRITE-UP}

\bigskip


\section{Introduction}
\label{sec:intro}

The method to solve the Skyrme Hartree-Fock-Bogolyubov equations in the
transformed harmonic oscillator basis was presented in \cite{[Sto05]}.
The present paper is a long write-up of the new version of the code HFBTHO.
This extended version contains a number of new capabilities such as the
breaking of reflection symmetry, the calculation of axial multipole moments,
multi-constraint calculations and the readjustment of the corresponding
Lagrange parameters using the cranking approximation of the RPA matrix, the
blocking prescription in odd-even and odd-odd nuclei, the finite-temperature
formalism, and generalized Skyrme-like energy functionals.

In addition to releasing a new version of the solver for general applications
in nuclear science, the goal of this paper is to establish a number of precise
benchmarks for nuclear structure calculations with Skyrme functionals. To
this end, we devote an entire section to comparing various calculations
performed with the spherical HOSPHE version 2.00 \cite{[Car13],[Car10]},
axially-deformed HFBTHO v{\codeversion}, and symmetry-unrestricted HFODD
version 2.56 \cite{[Sch13],[Sch12],[Dob09]} nuclear density functional theory
(DFT) solvers. Also, in order to facilitate the development of future versions
of HFBTHO as well as to enable deeper integration with the next releases of
HFODD, backward compatibility of input and output files has been broken between
the version 1.66 of \cite{[Sto05]} and the current version {\codeversion}.
Unless indicated otherwise, details about the methods presented in \cite{[Sto05]}
still apply.

In section \ref{sec:modifs}, we review the new capabilities of the code. In
section \ref{sec:benchmarks}, we present a number of numerical benchmarks
between HFBTHO and the aforementioned DFT solvers. Such benchmarks are very
important in view of the future development of these programs.


\section{Modifications introduced in version {\codeversion}}
\label{sec:modifs}

\setcounter{mysubsubsection}{0}

We present in this section the major new features added to the code between
version 1.66 and \codeversion. Minor improvements and bug fixes are not
discussed here, the full history of changes can be found in the source code.

\subsection{Modified Broyden Method}
\label{subsec:broyden}

In HFBTHO v{\codeversion}, the matrix elements of the HFB matrix are
updated at each iteration using the modified Broyden method, instead of the
traditional linear mixing of version 1.66. Details of the implementation,
results of convergence tests, and comparisons with alternative implementations
can be found in \cite{[Bar08]}.

\subsection{Axial multipole moments}
\label{subsec:moments}

In HFBTHO v{\codeversion}, the expectation value of axial multipole moments
$\hat{Q}_{l} \equiv \hat{Q}_{l0} = r^{l}Y_{l0}(\theta,\varphi)$ on the HFB
ground-state is computed for all moments up to $l_{\text{max}} = 8$. We
recall that in spherical coordinates, the multipole moment $\hat{Q}_{l}$
of order $l$ reads
\begin{equation}
\hat{Q}_{l}(r,\theta,\varphi) = r^{l}\sqrt{\frac{2l+1}{4\pi}} P_{l}(\cos\theta),
\end{equation}
where $P_{l}$ is the Legendre polynomial of order $l$ \cite{[Abr64]}. Spherical
and cylindrical coordinate systems are related through $r^{2} = \rho^{2} + z^{2}$
and $r\cos\theta = z$. Recurrence relations on Legendre polynomials give an
analytical expression for $\hat{Q}_{l}(r,z,\varphi)$ for $l=0,\dots,8$
\cite{[Abr64]}. Multipole moments can also be used as constraints. In this
case, the matrix elements of $\hat{Q}_{l}$ in the HO basis need to be computed. 
They are evaluated numerically on the Gauss-Laguerre and Gauss-Hermite nodes 
of integration used throughout the code \cite{[Sto05]}.

\subsection{Finite-temperature HFB method}
\label{subsec:temperature}

The code HFBTHO v{\codeversion} solves the finite temperature HFB (FT-HFB)
equations. The numerical implementation is similar to that of HFODD v2.49t
in \cite{[Sch12]}. Let us recall that the FT-HFB equations take the same
form as the HFB equations at $T=0$, only the one-body density matrix
and pairing tensor now depend on the Fermi-Dirac occupation $f_{\mu}$ of
quasi-particle states $\mu$. Assuming axial- and time-reversal symmetry,
all density matrices are real and read
\begin{equation}
\begin{array}{l}
\rho   = UfU^{T} + V(1-f)V^{T} \medskip\\
\kappa = UfV^{T} + V(1-f)U^{T},
\end{array}
\end{equation}
with $U,V$ the matrices of the Bogolyubov transformation. In HFBTHO, these
matrices are block-diagonal. As in HFODD, the Fermi level $\lambda$ is not
treated explicitly as the Lagrange parameter for the multipole operator
$\hat{Q}_{00}$ alongside other multipole moments $\hat{Q}_{lm}$. Instead,
it is determined directly at each iteration from the conservation
of particle number and is based on the BCS formula
\begin{equation}
N(\lambda) = \sum_{\mu} \left[ v_{\mu}(\lambda)^{2}
+ (u_{\mu}^{2}(\lambda)-v_{\mu}^{2}(\lambda))f_{\mu}(\lambda) \right].
\end{equation}
The BCS occupations are given by the traditional formulae
\begin{equation}
v_{\mu}^{2} = \frac{1}{2}\left[ 1 - \frac{\varepsilon_{\mu} - \lambda}{E_{\mu}^{\text{BCS}}} \right],
\ \ \ u_{\mu}^{2} = 1 - v_{\mu}^{2},
\end{equation}
with $E_{\mu}^{\text{BCS}} = \sqrt{(\varepsilon_{\mu} - \lambda)^{2} + \Delta_{\mu}^{2}}$
and $\varepsilon_{\mu}$ and $\Delta_{\mu}$ are the equivalent single-particle
energies and pairing gaps, see appendix B in \cite{[Dob84]}. The Fermi-Dirac
occupation factors are given by
\begin{equation}
f_{\mu}(\lambda) = \frac{1}{1 + e^{\beta E_{\mu}^{BCS}}}.
\end{equation}
When using the Newton-like method to solve the equation $N(\lambda)=N,Z$ for
each type of particle at $T>0$, one must now include the contribution
$\partial f_{\mu}/\partial\lambda$ in the derivative of the function
$N(\lambda)$.

\subsection{Linear constraints and the RPA method}
\label{subsec:rpa}

Multi-constraint calculations are possible in HFBTHO v{\codeversion}.
The code implements the linear constraint method, where the quantity
to be minimized is
\begin{equation}
E' = E - \sum_{a} \lambda_{a} \left( \langle \hat{Q}_{l_{a}} \rangle - Q_{l_{a}} \right),
\end{equation}
where $\hat{Q}_{l_{a}}$ is the multipole moment operator for the
constraint $a$ and $\lambda_{a}$ is the related Lagrange parameter.
Lagrange parameters are readjusted at each iteration according to
the procedure presented in \cite{[You09]} and also used in the latest
release of HFODD \cite{[Sch12]}. The philosophy of the method is to
associate the variation of the Lagrange parameters with a first-order
perturbation of the generalized density matrix.

As a reminder, we start with the variations $\delta\mathcal{R}$ of
the generalized density matrix, which induce variations of the HFB
matrix $\delta\mathcal{H}$ and of the Lagrange parameters
$\delta\gras{\lambda} = (\delta\lambda_{1},\dots,\delta\lambda_{N})$,
(up to first order). Neglecting the variations of the HFB matrix with
respect to the generalized density matrix is equivalent to working
at the so-called cranking approximation, and it reduces the HFB equation
with the perturbed quantities to
\begin{equation}
\left[ \delta\mathcal{R}, \mathcal{H}^{(0)} \right]
-\frac{1}{2} \sum_{a} \delta\lambda_{a}
\left[ \mathcal{R}^{(0)}, \mathcal{Q}_{l_{a}} \right] = 0,
\end{equation}
with $\mathcal{R}^{(0)}$ and $\mathcal{H}^{(0)}$, respectively, the
unperturbed generalized density matrix and HFB Hamiltonian,
$\delta\lambda_{a}$ the perturbation of the Lagrange parameter for the
constraint $a$, and $\mathcal{Q}_{l_{a}}$ the matrix of the constraint
in the doubled s.p.~basis. This equation gives the desired relation
between $\delta\mathcal{R}$ and $\delta\lambda$. The Lagrange parameter
can then be readjusted at each iteration by interpreting the deviation
$\delta Q_{l_{a}} = \langle \hat{Q}_{l_{a}} \rangle - Q_{l_{a}}$ from
the requested value $Q_{l_{a}}$ as caused by a variation of the
generalized density matrix $\delta\mathcal{R}$,
\begin{equation}
\delta Q_{l_{a}} = \frac{1}{2}\text{Tr} \left( \mathcal{Q}_{l_{a}}\delta\mathcal{R} \right).
\end{equation}
Knowing the deviation $\delta Q_{l_{a}}$, we obtain $\delta\mathcal{R}$,
and thereby deduce the $\delta\lambda_{a}$ needed to reproduce the
requested value. Calculations are performed in the q.p.~basis, since the
unperturbed generalized density and HFB matrix take a very simple form.
The computational cost of the method thus comes essentially from transforming
all relevant matrices into this basis. In HFBTHO, this operation can be
performed separately for each $\Omega-$block. The method can also be
extended to finite-temperature in a straightforward manner by using the
Fermi-Dirac occupation factors. Details of this extension are presented
elsewhere \cite{[Sch13]}.

\subsection{Quasi-particle blocking}
\label{subsec:blocking}

Odd-even and odd-odd nuclei can now be computed with HFBTHO v{\codeversion}
using the blocking of quasi-particle states \cite{[Sch10]}. Because
time-reversal symmetry is built into the code, the equal filling
approximation (EFA) has to be used \cite{[Per08]}. However, it was shown in
\cite{[Sch10]} that the EFA is an excellent approximation to exact
blocking. The identification of the blocking candidate is done using the
same technique as in HFODD \cite{[Dob09d]}: the mean-field Hamiltonian
$h$ is diagonalized at each iteration and provides a set of equivalent
single-particle states. Based on the Nilsson quantum numbers of the
requested blocked level provided in the input file, the code identifies
the index of the q.p.\ to be blocked by looking at the overlap between
the q.p.~wave-function (both lower and upper component separately) and
the s.p.~wave-function. The maximum overlap specifies the index of the
blocked q.p.

\subsection{Generalized energy density functionals}
\label{subsec:generalized}

The kernel of the HFBTHO solver has been rewritten to enable the use of
generalized Skyrme functionals that are not necessarily derived from an
effective pseudo-potential such as the Skyrme force. Generalized Skyrme
functionals are defined here as being the most general scalar, iso-scalar,
time-even functional $\mathcal{H}$ of the one-body local density matrix
$\rho(\gras{r})$ up to second-order in spatial derivatives of $\rho$
\cite{[Dob95],[Sto10]}. Assuming time-reversal symmetry, such functionals
thus take the form
\begin{equation}
\mathcal{H}_{t}[\rho] =
  C_{t}^{\rho\rho}[\rho] \rho_{t}^{2}
+ C_{t}^{\rho\tau}[\rho] \rho_{t}\tau_{t}
+ C_{t}^{J^{2}}[\rho] \gras{J}_{t}^{2}
+ C_{t}^{\rho\Delta\rho}[\rho] \rho_{t}\Delta\rho_{t}
+ C_{t}^{\rho\nabla J}[\rho] \rho_{t}\gras{\nabla}\cdot\gras{J}_{t},
\end{equation}
where $t$ stands for the isoscalar ($t=0$) or isovector ($t=1$) channel,
and $\tau_{t}$ and $\gras{J}_{t}$ are the kinetic energy and spin current
density in each channel. The terms $C_{t}^{uu'}[\rho]$ are (possibly
arbitrary) functions of the local isoscalar density $\rho_{0}(\gras{r})$.
Note that all commonly used Skyrme forces or functionals fall into this
category because of the phenomenological density-dependent term. Although
most Skyrme functionals have been fitted ``as a force", the recent
parameterizations UNEDF0 and UNEDF1 have looked at the problem more from a
functional perspective \cite{[Kor10],[Kor12]}. Microscopically-derived EDF
obtained, for example, from the density matrix expansion of effective
nuclear potentials, are less trivial examples of these generalized
functionals, since the density-dependence of the coupling constants can
be significant \cite{[Sto10]}.

In the current version, the code only implements 2$^{\text{nd}}$-order
generalized Skyrme functionals and it is left to the user to code more
advanced functionals. The subroutine {\tt calculate\_U\_parameters()} in
module {\tt UNEDF} provides a general template for such an implementation.
Required are the form of the energy functional and at least its first
partial derivatives with respect to the isoscalar $\rho_{0}$ and isovector
$\rho_{1}$ density matrices. Second-order partial derivatives are also
necessary to compute nuclear matter properties.

\subsection{Shared memory parallelism with OpenMP}
\label{subsec:omp}

To facilitate large-scale applications of the HFBTHO solver on leadership
class computers, the original source file has been split into a DFT solver
kernel and a calling program. In version {\codeversion}, we have also
parallelized a number of time-intensive routines using OpenMP pragmas. The
routine {\tt hfbdiag} diagonalizes the $\Omega-$blocks of the HFB matrix:
these diagonalizations are now done in parallel. The routine {\tt coulom}
computes the direct Coulomb potential $V_{\text{C}}(\gras{r},\gras{r}')$
at the first iteration: this step is carried out in parallel but saves
time at the first iteration only. The routine {\tt gamdel} reconstructs
the HFB matrix in configuration space for each $\Omega-$block by computing
on-the-fly the various one-dimensional integrals that define the matrix
elements: shared memory parallelism is implemented for the outermost loop
corresponding to the $\Omega-$blocks.

\begin{figure}[h]
\centering
\includegraphics[width=0.6\textwidth]{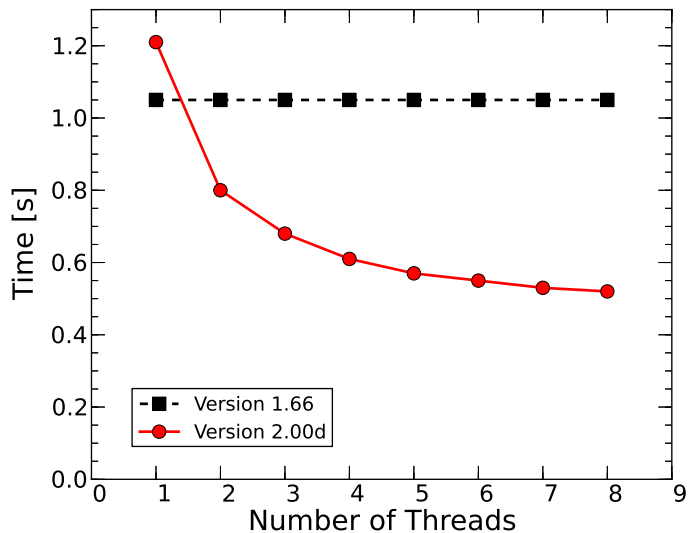}
\caption{Time per iteration of a HFB calculation in a full spherical basis
of $N_{\text{max}} = 20$ shells as a function of the number of threads,
see text for additional details.}
\label{fig:openmp}
\end{figure}

Figure \ref{fig:openmp} shows the performance improvement when using
multi-threading. The test was performed on an 8-core Intel Xeon E5-2670 at
2.6 GHz using the Intel Fortran compiler 13.0 and the MKL library 10.3.11
for $^{120}$Sn in a full spherical basis of  $N_{\text{max}} = 20$ shells,
with the SLy4 interaction and a cutoff of $E_{\text{cut}} = 60$ MeV for
the q.p. In version 1.66, the number of Gauss-Hermite and Gauss-Laguerre
points are hard-coded in the program and are set to $N_{\text{GH}} =
N_{\text{GL}} = 22$, and the number of Gauss-Legendre points is set to 
$N_{\text{GH}} = 30$. We used the same numbers in our test with the
version {\codeversion}.

We note that version {\codeversion} is slightly slower (per iteration)
than version 1.66 if only one thread is used. This additional overhead
comes from the calculation of the densities and fields required for
generalized Skyrme functionals, combined with the use of the Broyden
method, which uses additional linear algebra at each iteration. In 
general, it is difficult to compare directly the overall performance 
of the two versions of the code. In version 1.66, calculations at
$N_{\text{max}} > 14$ are warm-started automatically with a preliminary
calculation at $N_{\text{max}} = 14$. On the other hand, version
{\codeversion} implements the Broyden method, which reduces the number
of iterations significantly, see \cite{[Bar08]}. We emphasize
that HFBTHO makes use of BLAS and LAPACK routines and benefits from a
threaded implementation of these libraries. Nested parallelism must be
supported by the compiler.


\section{Benchmarks and Accuracy}
\label{sec:benchmarks}

There exist several comparisons of HFBTHO with other DFT solvers in the
literature in both even-even \cite{[Dob04a],[Pei08a]} and odd nuclei
\cite{[Sch09],[Sch10]}. In some cases, these benchmarks compare different
approaches to solving the HFB equations, in others the emphasis is put
on validation of the solver. Here, we want to gather in one place a
comprehensive set of validation and performance evaluations that can be
used as reference in later developments of the code.

\setcounter{mysubsubsection}{0}

\subsection{Benchmarks in spherical nuclei: $^{208}$Pb and $^{120}$Sn}
\label{subsec:test_spherical}

In spherical nuclei, HFBTHO was benchmarked against the spherical DFT solver
HOSPHE (version 2.00 of \cite{[Car13]}) and the symmetry-unrestricted DFT solver
HFODD (version 2.56 of \cite{[Sch13]}). We study the Hartree-Fock approximation
in $^{208}$Pb and the Hartree-Fock-Bogolyubov approximation with density-dependent
delta pairing forces in $^{120}$Sn.

\subsubsection{Hartree-Fock computation of $^{208}$Pb}
\label{subsubsec:test_pb}

\begin{table}[!ht]
\begin{center}
\begin{tabular}{c|cccc}
                                     &          HOSPHE               &           HFBTHO             &       HFODD                \\ \hline
                                     &                           \multicolumn{3}{c}{Without Coulomb}                             \\ \hline
$E_{\text{tot}}$ [MeV]               &-2445.93021{\bf\color{blue} 6} &-2445.93021{\bf\color{red} 6} &-2445.93021{\bf\color{db} 5}\\
$E^{\text{(n)}}_{\text{kin}}$ [MeV]  & 2614.806852                   & 2614.806852                  & 2614.806852                \\
$E^{\text{(p)}}_{\text{kin}}$ [MeV]  & 1438.160641                   & 1438.160641                  & 1438.160641                \\
$E_{\text{Skyrme}}$ [MeV]            &-6498.8977{\bf\color{blue} 08} &-6498.8977{\bf\color{red} 08} &-6498.8977{\bf\color{db} 06}\\
$E_{\text{SO}}$ [MeV]                & -109.091691                   & -109.091691                  & -109.091691                \\
$r^{\text{(n)}}_{\text{rms}}$ [fm]   &    5.519846                   &    5.519846                  &    5.519846                \\
$r^{\text{(p)}}_{\text{rms}}$ [fm]   &    5.2{\bf\color{blue} 49812} &    5.2{\bf\color{red} 50015} &    5.2{\bf\color{db} 50015}\\
\hline
                                     &                           \multicolumn{3}{c}{With Coulomb}                                \\ \hline
$E_{\text{tot}}$ [MeV]               &-1632.5914{\bf\color{blue} 19} &-1632.5914{\bf\color{red} 95} &-1632.5914{\bf\color{db} 54}\\
$E^{\text{(n)}}_{\text{kin}}$ [MeV]  & 2535.409{\bf\color{blue} 641} & 2535.409{\bf\color{red} 735} & 2535.409{\bf\color{db} 639}\\
$E^{\text{(p)}}_{\text{kin}}$ [MeV]  & 1340.663{\bf\color{blue} 301} & 1340.663{\bf\color{red} 408} & 1340.663{\bf\color{db} 312}\\
$E_{\text{Skyrme}}$ [MeV]            &-6306.660{\bf\color{blue} 514} &-6306.660{\bf\color{red} 740} &-6306.660{\bf\color{db} 527}\\
$E_{\text{SO}}$  [MeV]               &  -98.2933{\bf\color{blue} 31} &  -98.2933{\bf\color{red} 40} &  -98.2933{\bf\color{db} 31}\\
$E^{\text{(dir)}}_{\text{Cou}}$ [MeV]&  829.308{\bf\color{blue} 809} &  829.308{\bf\color{red} 760} &  829.308{\bf\color{db} 776}\\
$E^{\text{(exc)}}_{\text{Cou}}$ [MeV]&  -31.31265{\bf\color{blue} 6} &  -31.31265{\bf\color{red} 8} &  -31.31265{\bf\color{db} 6}\\
$r^{\text{(n)}}_{\text{rms}}$ [fm]   &    5.608237                   &    5.608237                  &    5.608237                \\
$r^{\text{(p)}}_{\text{rms}}$ [fm]   &    5.448{\bf\color{blue} 516} &    5.448{\bf\color{red} 711} &    5.448{\bf\color{db} 711}\\
\end{tabular}
\end{center}
\caption{Benchmark of the three solvers HOSPHE, HFBTHO and HFODD for a
spherical Hartree-Fock calculation in $^{208}$Pb with the SLy5 Skyrme
functional in a full spherical basis of $N_{\text{max}} = 16$ shells
with oscillator length $b = 2.0$ fm. See introduction of section
\ref{sec:benchmarks} for additional numerical information.}
\label{tab:pb208}
\end{table}

In table \ref{tab:pb208}, we present the results of the benchmarks between the
three solvers for the spherical HF point in $^{208}$Pb for the SLy5 Skyrme
functional of \cite{[Cha98]}. Calculations were performed in $N_{\text{max}}=16$
full spherical oscillator shells with a constant oscillator length of $b = 2.0$
fm. In a spherical basis, the oscillator length is related to the oscillator
frequency by
\begin{equation}
b = \sqrt{\frac{\hbar}{m\omega}}.
\end{equation}
In the codes HFODD and HOSPHE, the oscillator frequency is set via a multiplicative
factor $f$ such that $\omega = f\omega_{0}$, with $\omega_{0} = 41/A^{1/3}$. The
oscillator length is related to $f$ through
\begin{equation}
f = \frac{1}{b^{2}}\frac{\hbar^{2} c^{2}}{mc^{2}} / \frac{41}{A^{1/3}},
\end{equation}
with $mc^{2} = 938.90590$ MeV, $\hbar c = 197.328910$ MeV.fm. An oscillator length
of $b = 2.0$ fm thus corresponds to $f = 1.49831558$ in $^{208}$Pb. The number of
Gauss-Legendre points for the integration of the Coulomb potential was
$N_{\text{Leg}} = 80$, the number of Gauss-Hermite and Gauss-Laguerre integration
points was $N_{\text{GH}} = N_{\text{GL}} = 40$, and the Coulomb length scale was
$L=50$ fm, see also section \ref{subsec:test_coulomb} below for a detailed
discussion. The Skyrme energy is defined from HFBTHO outputs as
\begin{equation}
E^{\text{Skyrme}} = E^{\text{vol}} + E^{\text{surface}} + E^{\text{SO}} + E^{\text{tensor}}
\end{equation}

Without Coulomb potentials included, we note that the difference with HFODD is
not greater than 2 eV on energies ($E_{\text{Skyrme}}$), and the radii agree
up to at least the 6$^{\text{th}}$ digit. Comparisons with HOSPHE show the
difference in energies is less than 1 eV, while the proton radius differs by
0.0002 fm. This unexpected deviation may be caused by corrections related to 
the finite proton size which are first added to the proton radius and then
afterwards subtracted. Let us note that the kinetic energy contribution to the 
total energy is probably the most sensitive to the details of the numerical 
implementation. With the Coulomb potential included (both direct and exchange), 
the discrepancy on the total energy is of the order of 100 eV (see also section 
\ref{subsec:test_coulomb} below).

\subsubsection{Hartree-Fock-Bogolyubov computation of $^{120}$Sn with the Lipkin-Nogami prescription}
\label{subsubsec:test_sn}

\begin{table}[!ht]
\begin{center}
\begin{tabular}{c|ccc}
                                     &          HOSPHE               &           HFBTHO             &       HFODD                 \\ \hline
                                     &                           \multicolumn{3}{c}{Without Coulomb}                              \\ \hline
$E_{\text{tot}}$ [MeV]               &-1374.0876{\bf\color{blue} 78} &-1374.0876{\bf\color{red} 63} &-1374.0876{\bf\color{db} 31} \\
$E^{\text{(n)}}_{\text{kin}}$ [MeV]  & 1384.055{\bf\color{blue} 506} & 1384.055{\bf\color{red} 495} & 1384.055{\bf\color{db} 733} \\
$E^{\text{(p)}}_{\text{kin}}$ [MeV]  &  885.705{\bf\color{blue} 977} &  885.705{\bf\color{red} 972} &  885.705{\bf\color{db} 875} \\
$E_{\text{Skyrme}}$ [MeV]            &-3628.92{\bf\color{blue} 3716} &-3628.92{\bf\color{red} 3682} &-3628.92{\bf\color{db} 4197} \\
$E_{\text{SO}}$ [MeV]                &  -58.837{\bf\color{blue} 670} &  -58.837{\bf\color{red} 667} &  -58.837{\bf\color{db} 805} \\
$r^{\text{(n)}}_{\text{rms}}$ [fm]   &    4.678179                   &    4.678179                  &    4.678179                 \\
$r^{\text{(p)}}_{\text{rms}}$ [fm]   &    4.455{\bf\color{blue} 211} &    4.455{\bf\color{red} 761} &    4.455{\bf\color{db} 760} \\
$E_{\text{pair}}$ [MeV]              &  -12.64{\bf\color{blue} 6024} &  -12.64{\bf\color{red} 6023} &  -12.64{\bf\color{db} 5709} \\
$\Delta^{\text{(n)}}$ [MeV]          &    0.91007{\bf\color{blue} 1} &    0.91007{\bf\color{red} 1} &    0.91007{\bf\color{db} 2} \\
$\Delta^{\text{(p)}}$ [MeV]          &    0.5313{\bf\color{blue} 64} &    0.5313{\bf\color{red} 64} &    0.5313{\bf\color{db} 38} \\
$\lambda^{\text{(n)}}$ [MeV]         &   -7.3333{\bf\color{blue} 39} &   -7.3333{\bf\color{red} 39} &   -7.3333{\bf\color{db} 40} \\
$\lambda^{\text{(p)}}$ [MeV]         &  -21.40606{\bf\color{blue} 9} &  -21.40606{\bf\color{red} 9} &  -21.40606{\bf\color{db} 3} \\
$\lambda_{2}^{\text{(n)}}$ [MeV]     &    0.081422                   &    0.081422                  &    0.081422                 \\
$\lambda_{2}^{\text{(p)}}$ [MeV]     &    0.63330{\bf\color{blue} 5} &    0.63330{\bf\color{red} 5} &    0.63330{\bf\color{db} 9} \\ \hline
                                     &                           \multicolumn{3}{c}{With Coulomb}                                 \\ \hline
$E_{\text{tot}}$ [MeV]               &-1021.265{\bf\color{blue} 363} &-1021.265{\bf\color{red} 407} &-1021.265{\bf\color{db} 377} \\
$E^{\text{(n)}}_{\text{kin}}$ [MeV]  & 1345.2264{\bf\color{blue} 37} & 1345.2264{\bf\color{red} 97} & 1345.2264{\bf\color{db} 44} \\
$E^{\text{(p)}}_{\text{kin}}$ [MeV]  &  837.571{\bf\color{blue} 445} &  837.571{\bf\color{red} 518} &  837.571{\bf\color{db} 444} \\
$E_{\text{Skyrme}}$ [MeV]            &-3538.591{\bf\color{blue} 673} &-3538.591{\bf\color{red} 798} &-3538.591{\bf\color{db} 661} \\
$E_{\text{SO}}$ [MeV]                &  -48.652{\bf\color{blue} 094} &  -48.652{\bf\color{red} 102} &  -48.652{\bf\color{db} 100} \\
$E^{\text{(dir)}}_{\text{Cou}}$ [MeV]&  367.071{\bf\color{blue} 215} &  367.071{\bf\color{red} 165} &  367.071{\bf\color{db} 184} \\
$E^{\text{(exc)}}_{\text{Cou}}$ [MeV]&  -19.14010{\bf\color{blue} 3} &  -19.14010{\bf\color{red} 4} &  -19.14010{\bf\color{db} 3} \\
$r^{\text{(n)}}_{\text{rms}}$ [fm]   &    4.733892                   &    4.733892                  &    4.733892                 \\
$r^{\text{(p)}}_{\text{rms}}$ [fm]   &    4.585{\bf\color{blue} 076} &    4.585{\bf\color{red} 610} &    4.585{\bf\color{db} 609} \\
$E_{\text{pair}}$ [MeV]              &  -11.125231                   &  -11.125231                  &  -11.125231                 \\
$\Delta^{\text{(n)}}$ [MeV]          &    0.864875                   &    0.864875                  &    0.864875                 \\
$\Delta^{\text{(p)}}$ [MeV]          &    0.48123{\bf\color{blue} 6} &    0.48123{\bf\color{red} 7} &    0.48123{\bf\color{db} 6} \\
$\lambda^{\text{(n)}}$ [MeV]         &   -7.98957{\bf\color{blue} 3} &   -7.98957{\bf\color{red} 2} &   -7.98957{\bf\color{db} 3} \\
$\lambda^{\text{(p)}}$ [MeV]         &   -8.28670{\bf\color{blue} 3} &   -8.28670{\bf\color{red} 3} &   -8.28670{\bf\color{db} 4} \\
$\lambda_{2}^{\text{(n)}}$ [MeV]     &    0.100481                   &    0.100481                  &    0.100481                 \\
$\lambda_{2}^{\text{(p)}}$ [MeV]     &    0.67508{\bf\color{blue} 7} &    0.67508{\bf\color{red} 7} &    0.67508{\bf\color{db} 6} \\
\end{tabular}
\end{center}
\caption{Benchmark of the three solvers HOSPHE, HFBTHO and HFODD for a
spherical Hartree-Fock-Bogolyubov calculation in $^{120}$Sn with the
UNEDF0 functional (thus including the Lipkin-Nogami) prescription with
a spherical basis of $N_{\text{max}} = 16$ shells with oscillator scale
$b = 2.0$ fm ($f = 1.49831558$ in $^{120}$Sn). See introduction of section
\ref{sec:benchmarks} for additional numerical information.}
\label{tab:sn120}
\end{table}

In table \ref{tab:sn120}, we present the results of the benchmarks between the
three solvers for the spherical HFB point in $^{120}$Sn for the UNEDF0 Skyrme
functional of \cite{[Kor10]}. Calculations were performed with the same basis
and integration characteristics as in the previous section. The pairing channel
was parameterized by a density-dependent delta-pairing force with mixed volume
and surface features, of the general type
\begin{equation}
V_{\text{pair}}^{\text{(n,p)}}(\gras{r}) = V_{0}^{\text{(n,p)}}
\left( 1 - \frac{1}{2}\frac{\rho_{0}(\gras{r})}{\rho_{c}} \right) \delta(\gras{r} - \gras{r'}),
\label{eq:vpair}
\end{equation}
with $V_{0}^{\text{(n,p)}}$ the pairing strength for neutrons (n) and protons
(p), $\rho_{0}(\gras{r})$ the isoscalar local density, and $\rho_{c}$ the
saturation density, fixed at $\rho_{c} = 0.16$ fm$^{-3}$. Let us recall that
in the case of the UNEDF parameterizations, the pairing strengths should {\it
not} be adjusted by the user since they were fitted together with the Skyrme
coupling constants. Recommended values are, respectively, $V_{0}^{\text{(n)}}
= -170.374$ MeV and $V_{0}^{\text{(n)}} = -199.202$ MeV. Because of the
zero-range of the pairing force, a cutoff in the q.p.~space has to be
introduced, and we chose $E_{\text{cut}} = 60$ MeV in this example. When
compatibility with HFODD is required, this cutoff is sharp, namely all q.p.
with $E > E_{\text{cut}}$ are discarded from the calculation of the density.

\subsection{Benchmarks in even-even deformed nuclei: $^{240}$Pu}
\label{subsec:test_def}

Next, we present the benchmark of HFBTHO in deformed even-even nuclei
against HFODD. Accurate HFB calculations in deformed nuclei require the
use of a suitably deformed, or stretched, HO basis. Such a basis is
characterized by its oscillator frequencies,
$\omega_{x} \neq \omega_{y} \neq \omega_{z}$ in Cartesian coordinates,
and $\omega_{\perp} \neq \omega_{z}$ in cylindrical coordinates, as well
as by the total number of states retained. The goal of this section is
to compare basis truncation schemes between HFBTHO and HFODD in a
realistic case.

{\bf Stretched basis in HFBTHO - } It is determined by applying the general
prescription given in \cite{[Gir83]} to the particular case of an
axially-deformed prolate basis. Let us recall that the starting point is
an ellipsoid characterized by radii $R_{x}$, $R_{y}$ and $R_{z}$.
Introducing the spherical radius $R_{0}$ and the $(\beta,\gamma)$ Bohr
quadrupole deformation parameters, we have
\begin{equation}
\begin{array}{l}
\displaystyle
R_{x} = R_{0}\exp \left\{ \sqrt{\frac{5}{4\pi}} \beta\cos\left(\gamma - \frac{2\pi}{3} \right) \right\},
\medskip\\ 
\displaystyle
R_{y} = R_{0}\exp \left\{ \sqrt{\frac{5}{4\pi}} \beta\cos\left(\gamma + \frac{2\pi}{3} \right) \right\},
\medskip\\ 
\displaystyle
R_{z} = R_{0}\exp \left\{ \sqrt{\frac{5}{4\pi}} \beta\cos(\gamma) \right\}.
\end{array}
\label{eq:radii_tho}
\end{equation}
The deformation of the basis is characterized, equivalently, by the two
parameters $p$ and $q$ such that
\begin{equation}
q = \frac{b_{z}^{2}}{b_{x}^{2}},\ \ p = \frac{b_{y}^{2}}{b_{x}^{2}},
\end{equation}
with $b_{\mu}$ the oscillator length for coordinate $\mu$. It is then
assumed that
\begin{equation}
q = \frac{R_{z}}{R_{x}},\ \ p = \frac{R_{y}}{R_{x}}.
\end{equation}
In the special case of an axially-deformed basis ($\beta >0$, $\gamma = 0^{\text{o}}$),
we find
\begin{equation}
q = e^{-3\sqrt{\frac{5}{16\pi}}\beta},\ \ p=1.
\label{eq:15}
\end{equation}
From the volume conservation condition ($b_{\perp}^{2}b_{z}=b_{0}^{3}$),
Eq. (\ref{eq:15}) leads to
\begin{equation}
b_{\perp} = b_{0}q^{-1/6},\ \ b_{z} = b_{0}q^{+1/3}.
\label{eq:basis_THO}
\end{equation}
Given a ``spherical'' oscillator length $b_{0}$ and the deformation $\beta$
of the basis, the formula (\ref{eq:basis_THO}) uniquely defines the HO
lengths of the stretched basis.

{\bf Stretched basis in HFODD - } The starting point is a general nuclear
shape parameterized by a surface $\Sigma$ characterized by the deformation
parameters $\alpha_{\lambda\mu}$ through
\begin{equation}
R(\theta,\varphi) = R_{0}\,c(\alpha)\left[ 1 +
\sum_{\lambda=2}^{\lambda_{\text{max}}}\sum_{\mu=-\lambda}^{\lambda}
\alpha_{\lambda\mu}Y_{\lambda\mu}(\theta,\varphi) \right],
\label{eq:surface}
\end{equation}
where $R_{0} = r_{0}A^{1/3}$, $c(\alpha)$ is computed to ensure volume
conservation, $Y_{\lambda\mu}(\theta,\varphi)$ are the spherical harmonics,
and the $\alpha_{\lambda\mu}$ are the deformation parameters. The surface
defined by (\ref{eq:surface}) encloses a volume $V$ and the radius of
this ``ellipsoid'' along the direction $\mu$ (=$x,y,z$) is determined according
to
\begin{equation}
R_{\mu} \equiv \sqrt{\langle x_{\mu}^{2} \rangle} = \frac{1}{V} \int x_{\mu}^{2} d^{3}\gras{r}.
\label{eq:int}
\end{equation}
Frequencies of the HO along each Cartesian direction satisfy
$\omega_{0}^{3} = \omega_{x}\omega_{y}\omega_{z}$ with
\begin{equation}
\omega_{x} = \omega_{0} (R_{xz}R_{yz}), \ \ 
\omega_{y} = \omega_{0} (R_{xz}/R_{xy})^{-1/3}, \ \ 
\omega_{z} = \omega_{0} (R_{xy}R_{xz})^{1/3},
\end{equation}
with the geometrical ratios $R_{\mu\nu} = R_{\mu}/R_{\nu}$.

\begin{figure}[h]
\centering
\includegraphics[width=0.6\textwidth]{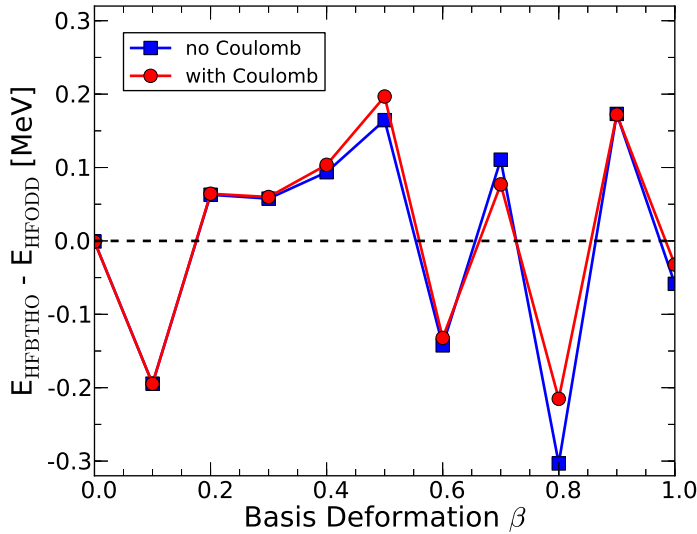}
\caption{Difference between HFBTHO and HFODD total energy for a very
deformed configuration in $^{240}$Pu (see details in text) as a function
of the axial deformation of the basis $\beta$.}
\label{fig:pu240}
\end{figure}

{\bf Discussion - } It is straightforward to see that in the particular
case of a prolate ellipsoid ($\beta >0$, $\gamma = 0^{\text{o}}$,
$R_{xy}=1/p$, $R_{xz}=R_{yz}=1/q$), both HFODD and HFBTHO prescriptions
to choose the oscillator frequencies are in principle identical. In
practice, however, the determination of the radii from Eq.(\ref{eq:int})
in HFODD produce small numerical deviations compared to the analytic
formula (\ref{eq:radii_tho}). This will induce systematic differences
between the HO frequencies computed in the two codes, which will in turn
alter the selection of the basis states. Figure \ref{fig:pu240} quantifies
this statement in an extreme case.

Figure \ref{fig:pu240} shows the numerical difference between the two
codes for the total energy in $^{240}$Pu computed for $N_{\text{max}}=16$
and $N_{\text{states}}=500$ as a function of the deformation $\beta$ of
the basis. Calculations are done with a spherical oscillator length
$b_{0} = 2.3$ fm ($f = 1.18829312$ for $^{240}$Pu), the SLy4
parameterization of the Skyrme functional, identical pairing strengths
of $V_{\text{pair}} = -300$ MeV for both protons and neutrons, and
quadrature precisions of $N_{\text{GL}}=N_{\text{GH}}=40$ and
$N_{\text{Leg}}=80$. The configuration chosen was obtained by putting a
constraint on the quadrupole moment $\langle \hat{Q}_{20}\rangle = 150$~b
and hexadecapole moment $\langle \hat{Q}_{40}\rangle = 30$ b$^{2}$;
expectation values of $\hat{Q}_{60}$ and $\hat{Q}_{80}$ vary with the
deformation of the basis.

For a configuration with such large deformations, a basis with only 500
states and $N_{\text{max}}=16$ is not sufficient to reach convergence.
In particular, important intruder orbitals are missing. As a result, all
physical observables depend quite significantly on basis parameters such
as its deformation or frequencies. It is therefore a good test-bench for
numerical comparisons and is an illustration of the worst-case scenario.

Two sets of results, with and without Coulomb potentials included, are
presented. In contrast to the much simpler cases of section
\ref{subsec:test_spherical}, the difference between the two codes reaches 
up to 300 keV, even without Coulomb terms. This discrepancy is entirely
attributable to the slightly different HO frequencies/lengths, the
impact of which is magnified by the large deformation of the requested
configuration combined with the relatively small size of the HO basis.
As an example, for $\beta = 0.7$, the oscillator lengths are
$b_{\perp} = 2.0803$ fm and $b_{z} = 2.8114$ fm in HFODD, to be compared
with $b_{\perp} = 2.0596$ fm and $b_{z} = 2.8682$ fm in HFBTHO. The
Coulomb term does not qualitatively change this picture. Most importantly,
{\em if HO lengths are manually enforced to be numerically identical in
the two codes}, or in the case of a spherical basis, the agreement
between the two sets of calculations without Coulomb goes back to the 1 eV
level as in the previous sections.

\subsection{Benchmark in deformed odd nuclei: $^{159}$Ba}
\label{subsec:test_blocking}

The new version of HFBTHO enables blocking calculations in odd-even or
odd-odd nuclei. Since by construction HFBTHO conserves time-reversal
symmetry, the blocking prescription is implemented in the equal filling
approximation, and the time-odd fields of the Skyrme functional are
identically zero. In \cite{[Sch10]}, a detailed comparison of blocking
calculations between the HFBTHO and HFODD solvers was presented for
the case of $^{121}$Sn, with a spherical HO basis and identical HO
oscillator scales. The goal of this section is to present a benchmark
result for an odd-even nucleus in a deformed basis. As in the previous
section, we do {\em not} manually enforce identical oscillator scales.
Instead, we use the same basis selection rules in their respective
implementations.

Calculations were performed in the nucleus $^{159}$Ba using $^{158}$Ba
as the even-even core, with the SLy4 Skyrme functional, a mixed
surface-volume pairing force with $V_{0} = -300$ MeV for both protons
and neutrons, and a q.p.~cutoff of $E_{\text{cut}} = 60$ MeV. The HO
basis was characterized by the oscillator length $b = 2.2$ fm
($f = 1.13221574$ in $^{159}$Ba), $N_{\text{max}}=16$ shells, a
deformation of $\beta=0.2$, and $N_{\text{States}}=500$. The number
of Gauss-Laguerre and Gauss-Hermite quadrature points was
$N_{\text{GL}} = N_{\text{GH}} = 40$. In HFODD calculations, time-odd
fields were zeroed. Results are presented in Table \ref{tab:blocking}.

\begin{table}[ht]
\begin{center}
\begin{tabular}{c|cc}
            & \multicolumn{2}{|c}{Total energy [MeV]}                        \\ \hline
   q.p.     &           HFBTHO              &       HFODD                    \\ \hline
 $[512]5/2$ & -1236.98{\bf\color{red} 2565} & -1236.98{\bf\color{blue} 5082} \\
 $[633]7/2$ & -1237.31{\bf\color{red} 7554} & -1237.31{\bf\color{blue} 6510} \\
 $[503]7/2$ & -1236.608{\bf\color{red} 320} & -1236.608{\bf\color{blue} 662} \\
 $[510]1/2$ & -1236.44{\bf\color{red} 5682} & -1236.44{\bf\color{blue} 8800} \\
 $[521]1/2$ & -1235.72{\bf\color{red} 4052} & -1235.72{\bf\color{blue} 6439} \\
 $[523]5/2$ & -1235.44{\bf\color{red} 5642} & -1235.44{\bf\color{blue} 9199} \\
 $[660]1/2$ & -1236.293{\bf\color{red} 614} & -1236.293{\bf\color{blue} 434} \\
 $[514]7/2$ & -1235.799{\bf\color{red} 601} & -1235.799{\bf\color{blue} 806} \\
 $[651]3/2$ & -1235.43{\bf\color{red} 3781} & -1235.43{\bf\color{blue} 4053} \\
\end{tabular}
\label{tab:blocking}
\end{center}
\caption{Results of blocking calculations in HFBTHO and HFODD in
$^{159}$Ba in a stretched HO basis with $\beta=0.2$ (see text for
more details).}
\end{table}

Numerical agreement is of the order of 1 keV, with maximum deviations
of up to 3.6 keV. Such an agreement is in line with the results shown in
the previous two sections. Since the nucleus is not as heavy as $^{240}$Pu
and the requested configuration is much less deformed than the one
considered in \ref{subsec:test_def}, basis truncation effects are
mitigated, and the small discrepancy between the calculated HO oscillator
scales does not have as drastic an effect as in the previous section.
Again, we note that if identical HO scales are manually enforced, the
numerical agreement is of the order of a few eV as shown in \cite{[Sch10]}.

\subsection{Transformed harmonic oscillator basis: $^{90}$Ni}
\label{subsec:test_tho}

One of the characteristic features of HFBTHO is the implementation of the
transformed harmonic oscillator (THO) basis. We recall that the THO basis
functions are generated by applying a local scale transformation (LST)
$f(\mathcal{R})$ to the HO single-particle basis functions. The LST
transforms every point $(\rho,z)$ by
\begin{equation}
\begin{array}{l}
\displaystyle
\rho \rightarrow \rho' = \rho\frac{f(\mathcal{R})}{\mathcal{R}}, \medskip\\
\displaystyle
z \rightarrow z' = z\frac{f(\mathcal{R})}{\mathcal{R}},
\end{array}
\end{equation}
with the scale $\mathcal{R} = \mathcal{R}(\rho,z)$ defined locally as
\begin{equation}
\mathcal{R} = \sqrt{\frac{\rho^{2}}{b_{\perp}^{2}} + \frac{z^{2}}{b_{z}^{2}}}.
\end{equation}
The LST function $f$ is chosen in such a way as to enforce the proper
asymptotic conditions (exponential decay) for the density, according to
the general procedure outlined in \cite{[Sto98],[Sto03]}. We refer to
\cite{[Sto05]} for the details of the implementation of the THO method
in HFBTHO.

\begin{table}[ht]
\begin{center}
\begin{tabular}{c|cc}
                                    &           HFBTHO              &       N. Michel's Code             \\ \hline
 $E_{\text{Skyrme}}$ [MeV]          & -2349.5{\bf\color{red} 47912} & -2349.5{\bf\color{blue} 11233} \\
 $E_{\text{SO}}$ [MeV]              &   -61.590{\bf\color{red} 852} &   -61.590{\bf\color{blue} 694} \\
 $E^{\text{(n)}}_{\text{kin}}$ [MeV]&  1190.9{\bf\color{red} 62911} &  1190.9{\bf\color{blue} 85716} \\
 $E^{\text{(n)}}_{\text{pair}}$[MeV]&   -58.{\bf\color{red} 593263} &   -58.{\bf\color{blue} 664803} \\
 $\Delta^{\text{(n)}}$ [MeV]        &     1.91{\bf\color{red} 5093} &     1.91{\bf\color{blue} 6030} \\
 $\lambda^{\text{(n)}}$ [MeV]       &    -0.19{\bf\color{red} 5904} &    -0.19{\bf\color{blue} 6557} \\
 $r_{\text{rms}}^{\text{(n)}}$ [fm] &     4.7173{\bf\color{red} 31} &     4.7173{\bf\color{blue} 75} \\
\end{tabular}
\label{tab:tho}
\end{center}
\caption{Results of THO calculations in HFBTHO and the spherical code of
\cite{[Mic08]} in $^{90}$Ni (see text for more details).}
\end{table}

The purpose of this section is to complete our collection of benchmarks 
by comparing the results obtained in the THO basis produced with HFBTHO with
an independent implementation of the method written by one of us (N. Michel)
and used in particular in \cite{[Sto08],[Mic08]}. This program assumes
spherical symmetry and has been developed independently: comparing the two
implementations is a particularly stringent test. To do it, we used HFBTHO
to generate the LST function $f(\mathcal{R})$ and its partial derivatives
on a spatial mesh $\mathcal{R}_{k}$ with $0\leq \mathcal{R} \leq 40$ fm by
steps of 0.02 fm. These functions were then read numerically by the
spherical code.

\begin{figure}[h]
\centering
\includegraphics[width=0.6\textwidth]{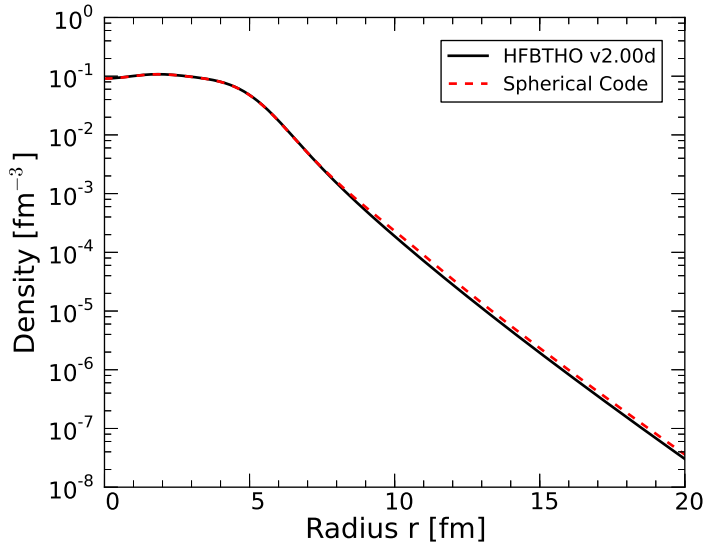}
\caption{Radial profile of the neutron density in $^{90}$Ni. Black plain line:
results from HFBTHO; red dashed line: results from the spherical code. }
\label{fig:densities}
\end{figure}

The test was carried out in the neutron-rich nucleus $^{90}$Ni, for the
SLy4 Skyrme functional and a pure surface pairing force characterized by
$V_{0}^{(n)} = V_{0}^{(p)} = -519.9$ MeV with a pairing cutoff of
$E_{\text{cut}} = 60$ MeV. Both the HO basis used to generate the THO basis
and the THO basis itself were spherical and contained $N_{\text{max}} = 20$
full shells. The oscillator length was fixed at $b = 2.0$ fm ($f = 1.13326033$
in $^{90}$Ni). The number of Gauss-Laguerre and Gauss-Hermite quadrature
points was $N_{\text{GL}} = N_{\text{GH}} = 40$. Both the direct and exchange
Coulomb terms were neglected in these tests. We present in Table \ref{tab:tho}
various quantities that are good indicators of potential numerical
discrepancies.

Overall, the agreement between the two implementations is very good. Indeed,
we recall that the matrix elements of the Hamiltonian in the THO basis depend
not only on the LST but also on its derivatives $\partial f/\partial\mathcal{R}$.
Using a numerically generated LST in the spherical code is, therefore, bound
to lead to systematic deviations. We show in figure \ref{fig:densities}
the radial profile of the corrected neutron density after the LST in both
HFBTHO and the spherical code. The tiny deviations beyond $r = 8$ fm are the
consequence of quantizing the LST in HFBTHO, and using this numerical function
in the spherical code instead of a native LST.

\subsection{Benchmark at finite temperature: $^{50}$Cr}
\label{subsec:test_temperature}

The new version of HFBTHO implements the finite-temperature HFB equations. 
Table \ref{tab:temperature} shows the comparison between HFBTHO and HFODD 
for a simple finite-temperature calculation in $^{50}$Cr. The characteristics 
of the test run (included with the submitted program) were the following: 
the calculation was performed in a full spherical basis of $N_{\text{max}} = 12$ 
shells, with an oscillator length of $b_{0} = 1.7622146$ fm (equivalent to 
$f = 1.2$), for the SLY4 interaction in the particle-hole channel and the 
standard surface-volume pairing force of Eq.(\ref{eq:vpair}) with 
$V_{0}^{(\text{n})} = V_{0}^{(\text{p})} = -300.0$ MeV and a cutoff of 
$E_{\text{cut}} = 60.0 $ MeV. The temperature was set at $T = 1.5$ MeV. We 
note that there is a bug in HFODD version 2.49t: the value of the entropy 
should be multiplied by a factor 2.

\begin{table}[ht]
\begin{center}
\begin{tabular}{c|cc}
                                    &           HFBTHO              &       HFODD                    \\ \hline
$E_{\text{tot}}$ [MeV]              & -423.4594{\bf\color{red} 51} & -423.4594{\bf\color{blue} 31} \\
$E^{\text{(n)}}_{\text{kin}}$ [MeV] &  461.5304{\bf\color{red} 78} &  461.5304{\bf\color{blue} 58} \\
$E^{\text{(p)}}_{\text{kin}}$ [MeV] &  402.309{\bf\color{red} 730} &  402.309{\bf\color{blue} 693} \\
$E_{\text{Skyrme}}$ [MeV]           &-1386.929{\bf\color{red} 749} &-1386.929{\bf\color{blue} 684} \\
$E_{\text{SO}}$ [MeV]               &  -36.60007{\bf\color{red} 7} &  -36.60007{\bf\color{blue} 1} \\
$r^{\text{(n)}}_{\text{rms}}$ [fm]  &    3.591964                  &    3.591964                   \\
$r^{\text{(p)}}_{\text{rms}}$ [fm]  &    3.594830                  &    3.594830                   \\
$\lambda^{\text{(n)}}$ [MeV]        &  -11.806442                  &  -11.806442                   \\
$\lambda^{\text{(p)}}$ [MeV]        &   -6.886468                  &   -6.886468                   \\
$S^{\text{(n)}}$ [MeV]              &    6.915625                  &    6.91562{\bf\color{blue} 5} \\
$S^{\text{(p)}}$ [MeV]              &    6.99515{\bf\color{red} 5} &    6.99515{\bf\color{blue} 8} \\
\end{tabular}
\label{tab:temperature}
\end{center}
\caption{Results of finite-temperature HFB calculations in HFBTHO and
HFODD in $^{50}$Cr in a spherical HO basis of $N_{\text{max}} = 12$ shells
(see text for more details).}
\end{table}

\subsection{Precision of the Coulomb term}
\label{subsec:test_coulomb}

In this section, we discuss in greater detail the precision of the direct term
of the Coulomb potential to the total energy. In HFBTHO, the direct term is
computed by the Gaussian substitution method. A less accurate method based on
the Laplacian substitution method is also available \cite{[Vau73]}.


\subsubsection{The Gaussian substitution method}

The direct term $V_{\text{Cou}}^{\text{(dir)}}(\gras{r})$ of the Coulomb
potential reads
\begin{equation}
V_{\text{Cou}}^{\text{(dir)}}(\gras{r})
= e^{2}\int d^{3}\gras{r}'\; \frac{\rho_{\text{p}}(\gras{r}')}{|\gras{r} - \gras{r}'|} ,
\end{equation}
with $\rho_{\text{p}}$ the proton density. In HFBTHO, the direct term is
computed by introducing the following expansion,
\begin{equation}
\frac{1}{|\gras{r} - \gras{r}'|}
= \frac{2}{\sqrt{\pi}}\int_{0}^{\infty} e^{-(\gras{r} - \gras{r}')^{2}/\mu^{2}}\frac{d\mu}{\mu^{2}}
= \frac{2}{\sqrt{\pi}}\int_{0}^{\infty} e^{-(\gras{r} - \gras{r}')^{2}a^{2}} da.
\end{equation}
Denoting
\begin{equation}
I_{a}(\gras{r}) = \int d^{3}\gras{r}'\; e^{-(\gras{r} - \gras{r}')^{2}a^{2}}\rho_{\text{p}}(\gras{r}'),
\end{equation}
we can write
\begin{equation}
V_{\text{Cou}}^{\text{(dir)}}(\gras{r}) = e^{2} \frac{2}{\sqrt{\pi}}\int_{0}^{\infty} I_{a}(\gras{r})da.
\label{eq:Vcou_1}
\end{equation}
The integral over the range $a$ can be performed by Gauss-Legendre quadrature if
we introduce the variable $0 \leq \xi < 1$ such that
\begin{equation}
a = \frac{1}{L} \frac{\xi}{\sqrt{1 -\xi^{2}}},
\label{eq:change_variable}
\end{equation}
with $L>0$ an arbitrary length scale. This leads to
\begin{equation}
V_{\text{Cou}}^{\text{(dir)}}(\gras{r})
= e^{2} \frac{2}{\sqrt{\pi}} \frac{1}{L}
\int_{0}^{1} \frac{I_{a(\xi)}(\gras{r})}{(1 - \xi^{2})^{3/2}}d\xi.
\label{eq:Vcou_2}
\end{equation}
The Coulomb direct energy is then given by
\begin{equation}
E_{\text{Cou}}
= \frac{1}{2}\int d^{3}\gras{r}\; V_{\text{Cou}}^{\text{(dir)}}(\gras{r})\rho_{\text{p}}(\gras{r})
= \int_{0}^{1} E_{\text{Cou}}(\xi)d\xi,
\end{equation}
with the integrand
\begin{equation}
E_{\text{Cou}}(\xi) =  e^{2} \frac{1}{\sqrt{\pi}} \frac{1}{L}
\int d^{3}\gras{r}
\frac{I_{a(\xi)}(\gras{r})\rho_{\text{p}}(\gras{r})}{(1 - \xi^{2})^{3/2}},
\label{eq:integrand}
\end{equation}

Let us note that the choice (\ref{eq:change_variable}) for the change
of variables is only a particular case of
\begin{equation}
a = \frac{1}{L} \frac{\xi}{(1 -\xi^{\alpha})^{1/\alpha}},
\end{equation}
with $\alpha$ any positive real number. In principle, $\alpha$ could be
tuned to maximize the convergence of the Coulomb energy with respect to
both the length scale $L$ and/or the number of points in the Legendre
quadrature, see below. In practice, the choice $\alpha = 2$ gives the
best compromise between accuracy and speed.


\subsubsection{Practical implementation in HFBTHO}

In practice, the integral (\ref{eq:Vcou_2}) is computed {\it numerically} by
introducing $N_{\text{Leg}}$ quadrature abscissae $\xi_{\ell}$ and weights
$w_{\ell}$,
\begin{equation}
V_{\text{Cou}}^{\text{(dir)}}(\gras{r})
= e^{2} \frac{2}{\sqrt{\pi}} \frac{1}{L}
\sum_{\ell=1}^{N_{\text{Leg}}} w_{\ell}\frac{I_{a(\xi_{\ell})}(\gras{r})}{(1 - \xi_{\ell}^{2})^{3/2}}.
\label{eq:Vcou_3}
\end{equation}

In HFBTHO, all integrals over $\gras{r} = (\rho,z,\varphi)$ are computed by
Gauss quadrature, with $N_{\text{GL}}$ Gauss-Laguerre points for the coordinate
$\eta=b_{\perp}^{2}\rho^{2}$ and $N_{\text{GH}}$ Gauss-Hermite points for the
coordinate $\xi=b_{z}z$ (see notations in \cite{[Sto05]}). Following the work
of Vautherin \cite{[Vau73]}, the code uses a general method known in electronic
structure theory as the pseudospectral representation of the HFB equations
\cite{[Fri85]}. While the HFB equations are solved in the HO basis, i.e. in Fock
(or spectral) space, the HF and pairing fields, as well as all expectations values
of observables, are computed directly on the quadrature grid, implying constant
transformation from/to Fock to/from coordinate space.

Following this philosophy, the calculation of the Coulomb field and energy
is somewhat accelerated by introducing the following matrix at the first
iteration,
\begin{equation}
V_{ki} = e^{2}\frac{2}{\sqrt{\pi}} \frac{1}{L}\; \omega_{i}
\sum_{\ell=1}^{N_{\text{Leg}}} w_{\ell}
\frac{e^{-(\gras{r}_{k} - \gras{r}'_{i})^{2}a(\xi_{\ell})^{2}}}{(1-\xi_{\ell}^{2})^{3/2}},
\label{eq:Vcou_disc}
\end{equation}
with $L$ the length scale mentioned in the previous section, $k,i$ compound
indexes running from 1 to $N_{\text{GL}}\times N_{\text{GH}}$, and $\omega_{i}$
the product of the weights for both types of quadrature,
$\omega_{i} \equiv w^{\text{GL}}w^{\text{GH}}$. With this notation, the Coulomb 
field on the grid is obtained at each iteration by vector multiplication 
$V_{k} = \sum_{i} V_{ki}\rho_{i}$, with $\rho_{i}$ the vector containing the 
proton density on the grid. The energy is then obtained by another vector 
multiplication $E_{\text{Cou}} = \sum_{k} \omega_{k}V_{k}\rho_{k}$.

It is important to bear in mind that the matrix $V_{ik}$ is a quantized
form of the true potential $1/|\gras{r} - \gras{r}'|$ on the quadrature
grid. When the original potential has a singularity, dealing with such
quantized representations generate a systematic error that can become
arbitrarily large near the singularity. This phenomenon is known as
``aliasing'' in electronic structure theory \cite{[Fri85]}. In principle,
the error should decrease as the grid becomes larger and larger (closer
to the exact integration). In any practical calculation, however, it will
always be non-zero.


\subsubsection{Numerical accuracy}

Mathematically, expressions (\ref{eq:Vcou_1}) and (\ref{eq:Vcou_2}) are
strictly equivalent. In particular, they do not depend on the length
scale $L$. However, the use of finite quadrature for {\it both} the
Gauss-Legendre integration of the $1/|\gras{r} - \gras{r}'|$ function
and the spatial integration over coordinates $\gras{r}$ and $\gras{r}'$
introduce an alias, as mentioned above. In our case, the practical
consequence of having an aliased integration is that the Coulomb energy
will depend, possibly in a significant way, on the length scale $L$ and
the number of Gauss-Legendre quadrature points $N_{\text{Leg}}$, but also on
the number of Gauss-Hermite and Gauss-Laguerre points $N_{\text{GH}}$ and
$N_{\text{GL}}$.

\begin{figure}[h]
\centering
\includegraphics[width=0.45\textwidth]{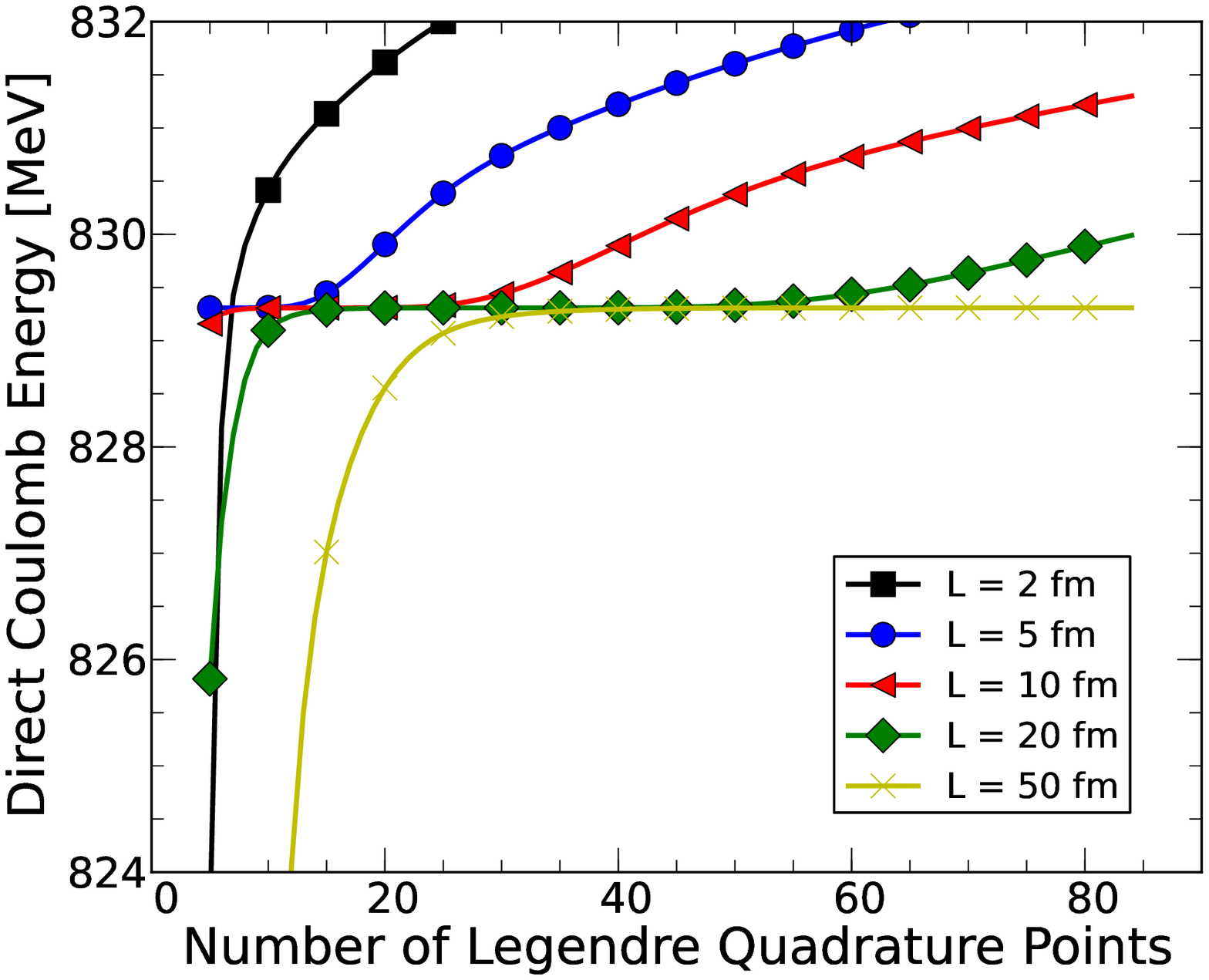}
\includegraphics[width=0.45\textwidth]{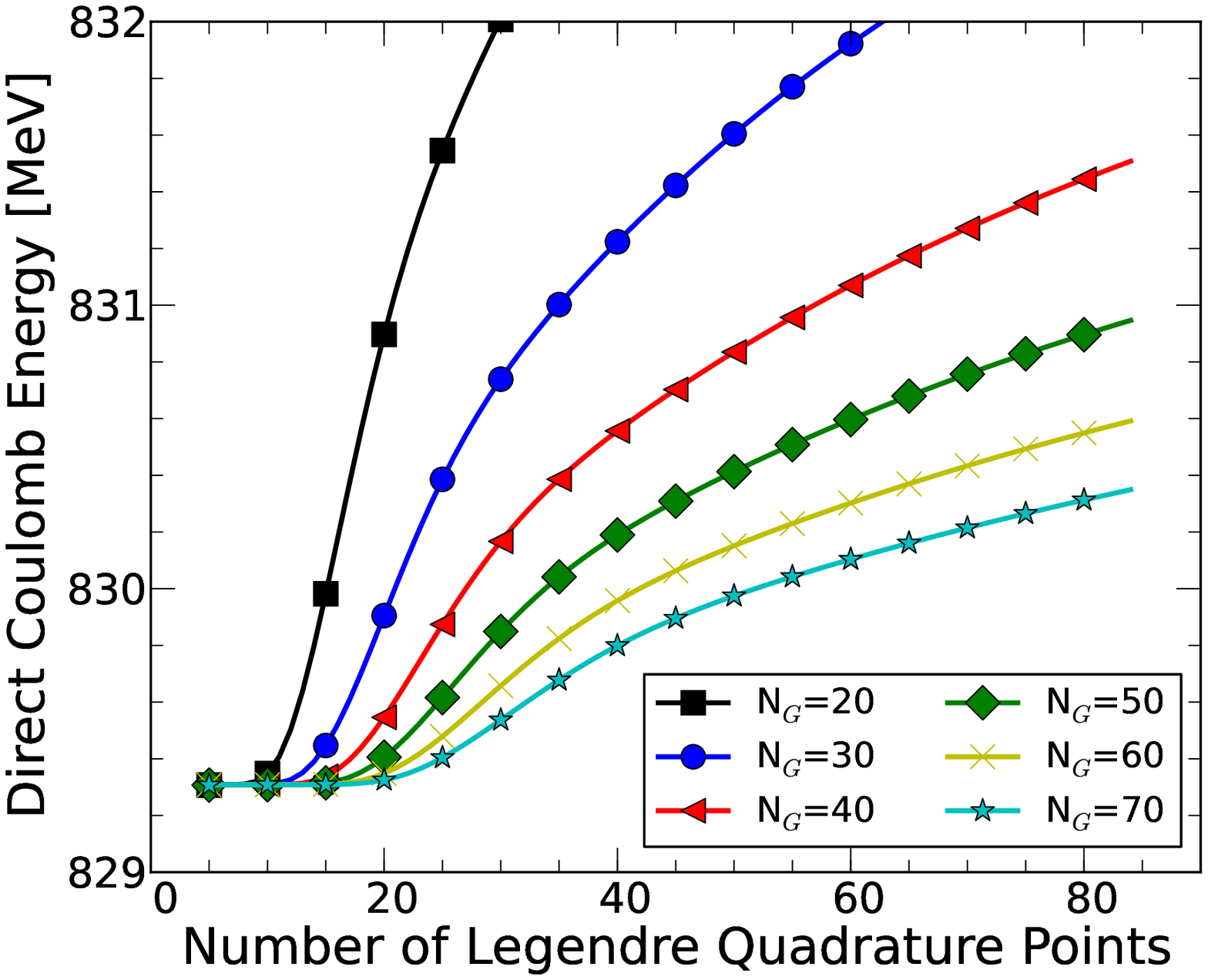}
\caption{Left: Direct Coulomb energy as a function of the number of
Gauss-Legendre integration points for different values of the length
scale $L$ (see text). Right: Same for $L=5$ fm and different values
of the Gauss-Hermite and Gauss-Laguerre quadrature points. All
calculations done in $^{208}$Pb for $N_{\text{max}}=16$ shells and
the SLy5 Skyrme functional.}
\label{fig:coulomb}
\end{figure}

In the left panel of figure \ref{fig:coulomb}, we show the direct
Coulomb energy in $^{208}$Pb as a function of the number of
Gauss-Legendre quadrature points for different length scales $L$.
Calculations were done in a full spherical HO basis with
$N_{\text{max}}=16$ oscillator shells and the SLy5 interaction with
$N_{\text{GH}} = N_{\text{GL}} = 30$. The dependence on $L$ is
clearly marked. In particular, there is no asymptotic convergence to
the true value of the Coulomb potential as the number of Legendre
integration points increases. Instead, one observes a plateau
condition, the range of which increases with $L$. In the right panel
of figure \ref{fig:coulomb}, we fix the length scale to $L=5$ fm, and
increase the precision of both Gauss-Hermite and Gauss-Laguerre
integrations (by convenience, we choose $N_{\text{GH}} = N_{\text{GL}}
\equiv N_{\text{G}}$). This clearly mitigates the dependence of the 
Coulomb energy on the Legendre quadrature.

\begin{figure}[h]
\centering
\includegraphics[width=0.45\textwidth]{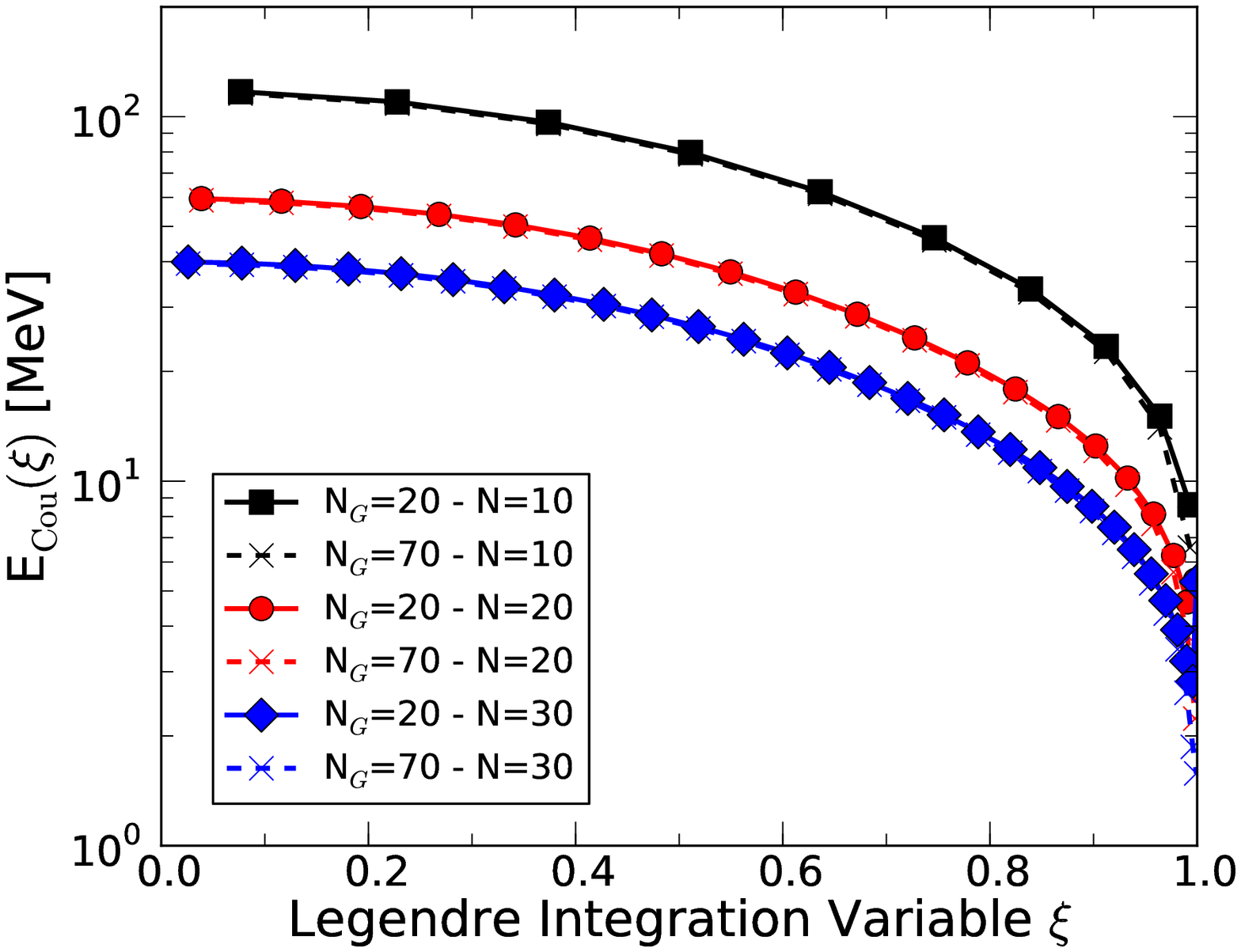}
\includegraphics[width=0.45\textwidth]{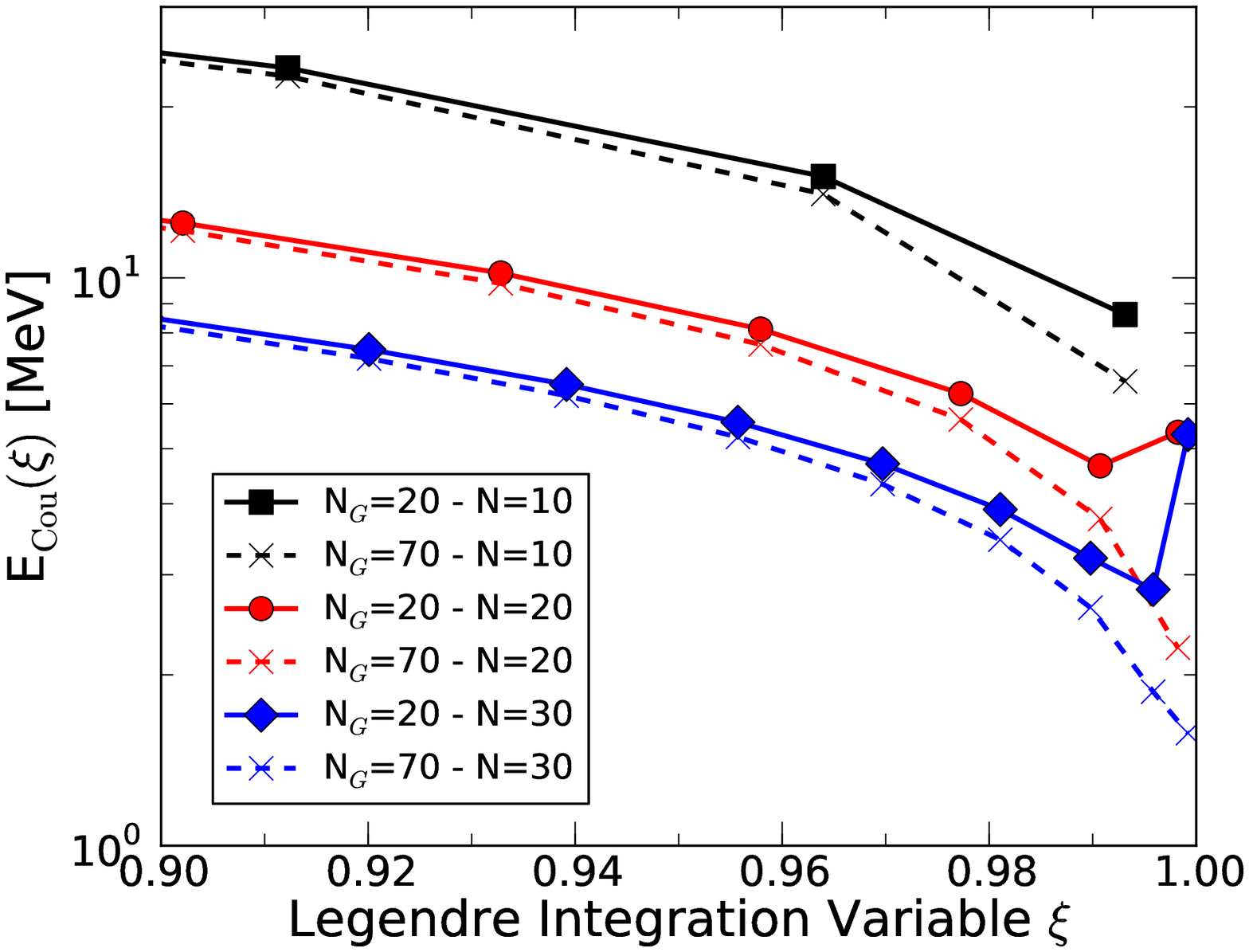}
\caption{Left: Integrand $E_{\text{Cou}}(\xi)$ of the Coulomb potential 
at convergence as function of the Gauss-Legendre integration variable 
$\xi$ (see text). Right: Close-up on the $\xi\in [0.9, 1.0[$ interval.}
\label{fig:coulomb1}
\end{figure}

It thus appears that the error is not really related to the Legendre
integration of Eq.(\ref{eq:Vcou_3}) itself. Instead, it seems to be a
consequence of using a finite quadrature for spatial integrations,
i.e. of dealing with a spurious alias. This effect can be visualized
in the behavior of the integrand (\ref{eq:integrand}). In figure
\ref{fig:coulomb1}, we show the integrand as a function of the variable
of integration $\xi$ for two types of quadrature meshes 
($N_{\text{G}} =20 $ or $N_{\text{G}} = 70$) and three different
numbers of Legendre integration points ($N_{\text{Leg}} = 10,20,30$). 
All calculations were done with a length scale $L=5$ fm. We recall that
the integrand should tend to 0 as $\xi\rightarrow 1$ (equivalent to
$a\rightarrow+\infty$). While the precision of the quadrature mesh does
not really play a role for most of the interval of variation of $\xi$,
we observe that for $\xi \rightarrow 1$, the function begins to bend up
for coarse quadrature grids (see right panel). This behavior is clearly
nonphysical and is the manifestation of the alias. It can be mitigated by
increasing the precision of the quadrature grid, as shown by the dashed
lines.

In HFBTHO, we have set $L=50$ fm and $N_{\text{Leg}} = 80$ as default
values. For calculations of ground-state properties, it is sufficient
to use the default values $N_{\text{GH}} = N_{\text{GL}} = 40$. For
calculations of very deformed configurations such as in fission, it
is recommended to increase the precision of Gauss integrations. In
future releases of the code, we will implement the calculation of both 
the direct and exchange Coulomb field in Fock space using Moshinsky 
transformations, which will eliminate all aliasing errors.


\section{Input data file}
\label{sec:input}

The input data file format has been entirely changed from version 1.66 to the
current version {\codeversion}. The number of additional features in the new
version was the reason to adopt a more flexible format for inputs.

\subsection{Sample input file}
\label{subsec:sample}

The new format uses Fortran namelist structure. An example is shown below,

\begin{verbatim}
&HFBTHO_GENERAL
 number_of_shells = 10, oscillator_length = -1.0, basis_deformation = 0.0,
 proton_number = 24, neutron_number = 26, type_of_calculation = 1 /
&HFBTHO_ITERATIONS
 number_iterations = 100, accuracy = 1.E-5, restart_file = -1 /
&HFBTHO_FUNCTIONAL
 functional = 'SLY4', add_initial_pairing = F, type_of_coulomb = 2 /
&HFBTHO_PAIRING
 user_pairing = F, vpair_n = -300.0, vpair_p = -300.0,
 pairing_cutoff = 60.0, pairing_feature = 0.5 /
&HFBTHO_CONSTRAINTS
 lambda_values = 1, 2, 3, 4, 5, 6, 7, 8,
 lambda_active = 0, 0, 0, 0, 0, 0, 0, 0,
 expectation_values = 0.0, 0.0, 0.0, 0.0, 0.0, 0.0, 0.0, 0.0 /
&HFBTHO_BLOCKING
 proton_blocking = 0, 0, 0, 0, 0, neutron_blocking = 0, 0, 0, 0, 0 /
&HFBTHO_PROJECTION
 switch_to_THO = 0, projection_is_on = 0,
 gauge_points = 1, delta_Z = 0, delta_N = 0 /
&HFBTHO_TEMPERATURE
 set_temperature = F, temperature = 0.0 /
&HFBTHO_DEBUG
 number_Gauss = 40, number_Laguerre = 40, number_Legendre = 80,
 compatibility_HFODD = F, number_states = 500, force_parity = T,
 print_time = 0 /
 \end{verbatim}

\subsection{Description of input data}
\label{subsec:description}

We now define the classes of input used in version {\codeversion}.

\key{HFBTHO\_GENERAL}

\noindent$\bullet$ {\tt number\_of\_shells = 10}: The principal number
of oscillator shells $N$. If the basis is spherical (see below), it is
made of the $N_{\text{states}} = (N+1)(N+2)(N+3)/6$ states
corresponding to $N$ full shells. If the basis is deformed, the code
searches for the lowest $N_{\text{states}}$, with possible intruder
contributions from up to the $N_{\text{max}} = 90$ HO shell. Default:
10.\\

\noindent$\bullet$ {\tt oscillator\_length = -1.0}: The oscillator
length in fm, denoted $b_{0}$ in this manuscript, corresponding
to the spherical basis. It is related to the HO frequency by
$b_{0} = \sqrt{\hbar/m\omega_{0}}$. If the basis is deformed, the code
uses the constant volume condition to define the $b_{z}$ and $b_{\perp}$
oscillator lengths such that $b_{0}^{3} = b_{z}b_{\perp}^{2}$. For negative
values of $b_{0}$, the code automatically sets $b_{0}$ by using
$\hbar\omega_{0} = 1.2\times 41/A^{1/3}$. Default: -1.0.\\

\noindent$\bullet$ {\tt basis\_deformation = 0.0}: The axial deformation
$\beta_{2}$ of the basis. Only axial quadrupole deformations are possible.
Negative values correspond to an oblate basis and are allowed. Default: 0.0.\\

\noindent$\bullet$ {\tt proton\_number = 24}: Number of protons for this
run. Only even values are allowed, see item {\tt proton\_blocking} under
keyword {\tt HFBTHO\_BLOCKING} for dealing with odd-proton nuclei. Default:
24. \\

\noindent$\bullet$ {\tt neutron\_number = 26}: Number of neutrons for this
run. Only even values are allowed, see item {\tt neutron\_blocking} under
keyword {\tt HFBTHO\_BLOCKING} for dealing with odd-neutron nuclei. Default:
26. \\

\noindent$\bullet$ {\tt type\_of\_calculation = 1}: Defines the type of
calculation to be performed for this run. If equal to 1, standard HFB
calculations will be performed. If equal to -1, the code will do HFB+LN,
where approximate particle-number projection is handled by the Lipkin-Nogami
prescription in the seniority pairing approximation following \cite{[Sch12]}.
Default: 1.

\key{HFBTHO\_ITERATIONS}

\noindent$\bullet$ {\tt number\_iterations = 100}: The maximum number of
iterations in the self-consistent loop. Default: 100. \\

\noindent$\bullet$ {\tt accuracy = 1.E-5}: Iterations are stopped when
the norm of the HFB matrix between two iterations,
$\text{max}|| \mathcal{H}^{(n)} - \mathcal{H}^{(n-1)}||$, is lower than 
{\tt accuracy}, or the number of iterations has exceeded 
{\tt number\_iterations}. Default: 1.E-5.\\

\noindent$\bullet$ {\tt restart\_file = -1}: This key can take the 
values $\pm 1, \pm 2, \pm 3$. If it is negative, calculations will be restarted 
from an existing solution stored in a HBFTHO compatible binary file. The name 
of this file will always take the form 
{\tt [shape][neutron\_number]\_[proton\_number].[extension]}, where {\tt [shape]} 
is `{\tt s}' for the value $\pm 1$, `{\tt p}' for the value $\pm 2$ and `{\tt o}' 
for the value $\pm 3$, and {\tt [extension]} is either `{\tt hel}' for regular HO 
calculations or `{\tt tel}' for THO calculations. If the value of the key is 
positive, calculations will be started from scratch by solving the Schr{\"o}dinger 
equation for a Woods-Saxon potential with (possibly) an axial deformation $\beta_{2}$ 
defined by the value of the constraint on $Q_{2}$, see below. Default: -1.

\key{HFBTHO\_FUNCTIONAL}

\noindent$\bullet$ {\tt functional = `SLY4'}: This key with 4 letters
indicates the Skyrme functional to be used. Possible values are: {\tt `SIII'},
{\tt `SKM*'}, {\tt `SKP'}, {\tt `SLY4'}, {\tt `SLY5'}, {\tt `SLY6'}, {\tt `SLY7'},
{\tt `SKI3'}, {\tt `SKO'}, {\tt `SKX'}, {\tt `HFB9'}, {\tt `UNE0'},
{\tt `UNE1'}. Default: {\tt `SLY4'}.\\

\noindent$\bullet$ {\tt add\_initial\_pairing = F}: In restart mode (see
{\tt restart\_file} ), this boolean variable decides if a small number will be added to
all pairing matrix elements. This option can be useful to ensure that
pairing correlations remain non-zero even when restarting from a nucleus
where they have collapsed, such as a doubly-magic nucleus. Default: F. \\

\noindent$\bullet$ {\tt type\_of\_coulomb = 2}: Chooses how the Coulomb
potential is treated. If 0, both the direct and exchange terms are
neglected. If 1, only the direct Coulomb potential is included in the
calculation. If 2, both the direct and exchange Coulomb potentials are
included, the exchange term being treated in the Slater approximation.
Default: 2.\\

\key{HFBTHO\_PAIRING}

\noindent$\bullet$ {\tt user\_pairing = T}: When this keyword is set to
{\tt T}, some characteristics of the pairing interaction can be set by
the user. It is always assumed that the pairing force reads
\begin{equation}
V_{\text{pair}}^{\text{n,p}}(\gras{r}) =
V_{0}^{\text{n,p}}\left( 1 - \alpha \frac{\rho(\gras{r})}{\rho_{c}} \right)
\delta(\gras{r}-\gras{r'}).
\label{eq:Vpair}
\end{equation}
Parameters that can be adjusted are the value of the pairing strength
for protons and neutrons $V_{0}^{\text{n,p}}$ (which can be different),
the cutoff in energies defining the q.p.\ entering the calculation of
the densities, and the type of pairing force defined by the parameter
$\alpha$. When this keyword is set to {\tt F}, a pre-defined pairing
force is used for each Skyrme functional. Default: F.\\

\noindent$\bullet$ {\tt vpair\_n = -300.0}: The value of the pairing
strength (in MeV) for neutrons $V_{0}^{\text{n}}$ in Eq.(\ref{eq:Vpair}).
Default: depends on the Skyrme force.\\

\noindent$\bullet$ {\tt vpair\_p = -300.0}: The value of the pairing
strength (in MeV) for protons $V_{0}^{\text{p}}$ in Eq.(\ref{eq:Vpair}).
Default: depends on the Skyrme force.\\

\noindent$\bullet$ {\tt pairing\_cutoff = 60.0}: The energy cutoff
(in MeV) in q.p.~space: all q.p.\ with energy lower than the cutoff
are taken into account in the calculation of the densities. Default:
60.0 MeV.\\

\noindent$\bullet$ {\tt pairing\_feature = 0.5}: The factor $\alpha$
in Eq.(\ref{eq:Vpair}). This parameter enables one to tune the properties
of the pairing force: If equal to 0, the pairing force has pure volume
character and does not depend on the isoscalar density; if set to 1,
the pairing force is only active at the surface, since in the interior,
$\rho(\gras{r}) \approx \rho_{c}$; if set to 0.5, the pairing force
has mixed volume-surface characteristics. Only values between 0 and 1
are possible. Default: 0.5.\\

\key{HFBTHO\_CONSTRAINTS}

\noindent$\bullet$ {\tt lambda\_values = 1, 2, 3, 4, 5, 6, 7, 8}: This
series of 8 integers define the multipolarity of the multipole moment
constraints. It is informational only and is not meant to be changed. \\

\noindent$\bullet$ {\tt lambda\_active = 0, 0, 0, 0, 0, 0, 0, 0}: This
line defines which of the multipole moments operator $\hat{Q}_{l}$, for
$l=1,\dots,8$, will be used as constraints. When 0, the corresponding
multipole is not used as constraint. When 1 it is used, and the resulting 
constrained HFB calculation is initialized from the diagonalization of 
the Woods-Saxon potential with the basis deformations. The user can also 
set this key to -1, which triggers the kickoff mode: the code first 
performs up to 10 iterations with the constraints specified by the keyword 
{\tt expectation\_values} below, then releases all constraints so as to 
reach the nearest unconstrained solution. Default:
{\tt (/ 0, 0, 0, 0, 0, 0, 0, 0 /)} (unconstrained calculations).\\

\noindent$\bullet$ {\tt expectation\_values = 0.0, 0.0, 0.0, 0.0, 0.0,
0.0, 0.0, 0.0}: This line complements the preceding one by specifying
the value of the constraint for each multipolarity $l$. Internally, the
units for the multipole moment of order $l$ are 10$^{l}\times$fm$^{l}$.
Example: In order to obtain a constraint value of $Q_{3} = 5$ b$^{3/2} =
5000$ fm$^{3}$, the third number must be set to 5.0. Default:
{\tt (/ 0, 0, 0, 0, 0, 0, 0, 0 /)}.\\

\key{HFBTHO\_BLOCKING}

\noindent$\bullet$ {\tt proton\_blocking = 0, 0, 0, 0, 0}: This group
of 5 integers defines the blocking configuration for protons. It takes
the form $2\Omega, \pi, N, n_{z}, n_{r}$, where $[N,n_{z},n_{r}]\Omega^{\pi}$
is the traditional Nilsson label. Recall that with time-reversal symmetry,
states $+\Omega$ and $-\Omega$ are degenerate, and HFBTHO only considers
states with positive values of $\Omega$ by default: the sign of $2\Omega$
given above is not related to the actual value of $\Omega$, but to the
nucleus in which the blocking is performed. Specifically,
\begin{itemize}
\item If $2\Omega=0$, the entire group is disregarded (no blocking).
\item If $2\Omega>0$, blocking is carried out in the nucleus with
$Z+1$ protons, where $Z$ is the value given by the flag {\tt proton\_number}.
In practice, it means the resulting HFB solution corresponds to the $(Z+1,N)$
nucleus.
\item If $2\Omega<0$, blocking is carried out in the nucleus with
$Z-1$ protons, where $Z$ is the value given by the flag {\tt proton\_number}.
In practice, it means the HFB solution corresponds to the $(Z-1,N)$
nucleus.
\end{itemize}
Additionally, the user may request all blocking configurations within
2 MeV of the Fermi level in the even-even core to be computed. This
automatization is activated by setting the parity $\pi$ to 0 instead
of $\pm 1$. For example, the line {\tt 1, 0, 0, 0, 0} would compute
all blocking configurations in the $(Z+1,N)$ nucleus, while the line
{\tt -7, -1, 3, 0, 3} would yield the configuration [303]7/2$^{-}$
in the $(Z-1,N)$ nucleus. Refer to the examples included with the program
for a practical application. Default: {\tt (/ 0, 0, 0, 0, 0 /)}. \\

\noindent$\bullet$ {\tt neutron\_blocking = 0, 0, 0, 0, 0}: This group
of 5 integers defines the blocking configuration for neutrons. It obeys
the same rules as for the protons. Default: {\tt (/ 0, 0, 0, 0, 0 /)}.\\

\key{HFBTHO\_PROJECTION}

\noindent$\bullet$ {\tt switch\_to\_THO = 0}: This switch controls the
use of the transformed harmonic oscillator basis. If equal to 0, then
the traditional HO basis is used; if equal to -1, then the code first
performs a calculation in the HO basis before automatically restarting
the calculation in the THO basis after the local scale transformation
has been determined; if 1, the code runs the calculation in the THO
basis only. Note that the use of the THO option requires a large enough
basis, typically with at least $N_{\text{max}} = 20$. Default: 0.\\

\noindent$\bullet$ {\tt projection\_is\_on = 0}: Particle number
projection (after variation) is activated by switching this integer
to 1. Default: 0.\\

\noindent$\bullet$ {\tt gauge\_points = 1}: The implementation of
particle number projection is based on the discretization of the
integration interval over the gauge angle. The number of gauge points
is given here. Default: 1.\\

\noindent$\bullet$ {\tt delta\_Z = 0}: If particle projection is on,
HFB results will be projected on $Z+\delta Z$, where $Z$ is the actual
number of protons in the nucleus and $\delta Z$ is specified here.
Default: 0\\

\noindent$\bullet$ {\tt delta\_N = 0}: If particle projection is on,
HFB results will be projected on $N+\delta N$, where $N$ is the actual
number of neutrons in the nucleus and $\delta N$ is specified here.
Default: 0\\

\key{HFBTHO\_TEMPERATURE}

\noindent$\bullet$ {\tt set\_temperature = F}: For finite-temperature
HFB calculations, {\tt set\_temperature} must be set to T. Default: F.\\

\noindent$\bullet$ {\tt temperature = 0.0}: In finite-temperature
HFB calculations, the value of the nuclear temperature is given here,
in MeV. If {\tt set\_temperature = F}, but the nuclear temperature is
positive, the code overwrites the flag {\tt set\_temperature}. Default:
0.0.\\

\key{HFBTHO\_DEBUG}

\noindent$\bullet$ {\tt number\_Gauss = 40}: Number of Gauss-Hermite
integration points for integrations along the z-axis (elongation axis).
Default: 40 (conserved parity), 80 (broken parity).\\

\noindent$\bullet$ {\tt number\_Laguerre = 40}: Number of Gauss-Laguerre
integration points for integrations along the perpendicular axis. Default:
40.\\

\noindent$\bullet$ {\tt number\_Legendre = 80}: Number of Gauss-Legendre
integration points for the calculation of the direct Coulomb potential,
see section 3.8 of \cite{[Sto05]} and section \ref{subsec:test_coulomb} in
this manuscript. If this number is negative, the Laplacian substitution
method is used instead of the Gaussian substitution method, see
\cite{[Vau73]}. Default: 80.\\

\noindent$\bullet$ {\tt compatibility\_HFODD = F}: This boolean flag
enforces the same HO basis as in HFODD. In practice, it is only meaningful
in deformed nuclei. Default: F.\\

\noindent$\bullet$ {\tt number\_states = 500}: When compatibility with
HFODD conventions is enforced, this parameter gives the total number of
states in the basis. Default: Inactive.\\

\noindent$\bullet$ {\tt force\_parity = T}: This boolean flag enforces the
conservation or breaking of parity depending on the multipolarity of
the multipole moments used as constraints. Default: T.\\

\noindent$\bullet$ {\tt print\_time = 0}: If 1, the time taken by
some of the major routines will be printed in the output. Default: 0.


\section{Program HFBTHO v{\codeversion}}
\label{sec:program}

The program HFBTHO comes in the form of two files:
\begin{itemize}
\item \tv{hfbtho\_200d.f90} - Main file containing the self-contained
HFBTHO solver. This file contains several Fortran modules, see below.
\item \tv{main\_200d.f90} - Calling program.
\end{itemize}
The programming language of most of the code is Fortran 95, while legacy code
is still written, in part or totally, in Fortran 90 and Fortran 77. The code
\pr{HFBTHO} requires an implementation of the BLAS and LAPACK libraries to
function correctly. Shared memory parallelism is available.

\subsection{Fortran Source Files}
\label{subsec:source}

The main file \tv{hfbtho\_200d.f90} contains the following Fortran
modules:
\begin{itemize}
\item {\tt HFBTHO\_VERSION}: informational module only containing the change
log;
\item {\tt HFBTHO\_utilities}: definition of integer and real number types;
\item {\tt linear\_algebra}: collection of various routines dealing with
interpolation;
\item {\tt UNEDF}: module computing the Skyrme-like energy density and the
corresponding Hartree-Fock fields at a given density $\rho$;
\item {\tt HFBTHO}: module storing all public variables used throughout the
code;
\item {\tt HFBTHO\_gauss}: collection of routines and functions dealing with
the integration meshes (contains several Fortran 77 legacy routines);
\item {\tt HFBTHO\_THO}: module in charge of the THO transformation;
\item {\tt EllipticIntegral}: module that provides the elliptic integral of
the second kind;
\item {\tt bessik}: module that provides the modified Bessel function
of integer order.
\end{itemize}
The rest of the routines are not stacked into a module.

\subsection{Compilation}
\label{subsec:compilation}

The program is shipped with a Makefile that is preset for a number of
Fortran compilers. The user should choose the compiler and set the path
for the BLAS and LAPACK libraries. To compile, type: ``{\tt make}".

\subsection{Code execution}
\label{subsec:execution}

Assuming an executable named {\tt main} and a Linux system, execution is
started by typing

\begin{center}
``{\tt ./main < /dev/null >\& main.out }"
\end{center}

The program will attempt to read the file named {\tt hfbtho\_NAMELIST.dat}
in the current directory. The user is in charge of assuring this file is
present and readable. The code will automatically generate a binary file
of the form {\tt [shape][neutron\_number]\_[proton\_number].[extension]}
where:
\begin{itemize}
\item {\tt [shape]} is one of the letters `s', `p', `o', which refers to
spherical, prolate or oblate shape respectively. The choice of this letter
is left to the user through the keyword {\tt restart\_mode}. This format
remains for backward compatibility;
\item {\tt [neutron\_number]} is a 3-integer number giving the neutron number
(left-padding with zero if necessary);
\item {\tt [proton\_number]} is a 3-integer number giving the proton number
(left-padding with zero if necessary);
\item {\tt [extension]} is either `hel' (normal HO run) or `tel' (THO run).
\end{itemize}


\section{Acknowledgments}
\label{sec:acknowledgments}

\bigskip
Discussions with R. Parrish are very warmly acknowledged.
Support for this work was partly provided through Scientific Discovery 
through Advanced Computing (SciDAC) program funded by U.S. Department 
of Energy, Office of Science, Advanced Scientific Computing Research 
and Nuclear Physics; by the the Academy of Finland under the Centre 
of Excellence Programme 2012-2017 (Nuclear and Accelerator Based 
Physics Programme at JYFL) and FIDIPRO programme, the U.S.\ 
Department of Energy grant Nos.~DE-FC02-09ER41583, DE-FC02-07ER41457, 
DE-FG02-96ER40963 (University of Tennessee), and DE-AC02006CH11357 
(Argonne National Laboratory). It was partly performed under the 
auspices of the US Department of Energy by the Lawrence Livermore 
National Laboratory under Contract DE-AC52-07NA27344 (code release 
number: LLNL-CODE-573953, document release number: LLNL-JRNL-587360). 
Funding was also provided by the United States Department of Energy 
Office of Science, Nuclear Physics Program pursuant to Contract 
DE-AC52-07NA27344 Clause B-9999, Clause H-9999 and the American 
Recovery and Reinvestment Act, Pub. L. 111-5. An award of computer 
time was provided by the Innovative and Novel Computational Impact 
on Theory and Experiment (INCITE) program. This research used 
resources of the Oak Ridge Leadership Computing Facility located 
in the Oak Ridge National Laboratory, which is supported by the 
Office of Science of the Department of Energy under Contract 
DE-AC05-00OR22725. It also used resources of the National Energy 
Research Scientific Computing Center, which is supported by the 
Office of Science of the U.S. Department of Energy under Contract 
No. DE-AC02-05CH11231. We also acknowledge ``Fusion,''a 320-node 
cluster operated by the Laboratory Computing Resource Center at 
Argonne National Laboratory, and the CSC-IT Center for Science 
Ltd, Finland for the allocation of computational resources.


\bibliographystyle{cpc}

\begin{thebibliography}{10}

\bibitem{[Sto05]}
{M.V.~Stoitsov, J.~Dobaczewski, W.~Nazarewicz, and P.~Ring, Comput. Phys.
  Commun. {\bf 167}, 43 (2005).}

\bibitem{[Car13]}
{B.G. Carlsson, J. Toivanen, J. Dobaczewski, P. Vesely, Y. Gao, {\it In preparation}, (2013).}

\bibitem{[Car10]}
{B.G. Carlsson, J. Dobaczewski, J. Toivanen, P. Vesel\'{y}, Comput. Phys.
  Commun. {\bf 181}, 1641 (2010).}

\bibitem{[Sch13]}
{N. Schunck, J. Dobaczewski, R. Parrish, W. Satu{\l}a, {\it In preparation}, (2013) .}

\bibitem{[Sch12]}
{N. Schunck, J. Dobaczewski, J. McDonnell, W. Satu{\l}a, J.A. Sheikh, A.
  Staszczak, M.V. Stoitsov, and P. Toivanen, Comput. Phys. Commun. {\bf 183}
  166 (2012).}

\bibitem{[Dob09]}
{J. Dobaczewski, W. Satu{\l}a, B.G. Carlsson, J. Engel, P. Olbratowski, P.
  Powa{\l}owski, M. Sadziak, J. Sarich, N. Schunck, A. Staszczak, M.V.
  Stoitsov, M. Zalewski, and H. Zdu\'nczuk, HFODD (v2.40h) User's Guide:
  arXiv:0909.3626 (2009)}.

\bibitem{[Bar08]}
{A. Baran, A. Bulgac, M. McNeil Forbes, G. Hagen, W. Nazarewicz, N. Schunck,
  and M.V. Stoitsov, Phys. Rev. C {\bf 78}, 014318 (2008).}

\bibitem{[Abr64]}
{M. Abramowitz and I.A. Stegun, {\it Handbook of mathematical functions},
  (Dover Publications, 1964).}

\bibitem{[Dob84]}
{J. Dobaczewski, H. Flocard and J. Treiner, Nucl. Phys. A {\bf 422}, 103
  (1984).}

\bibitem{[You09]}
{W. Younes and D. Gogny, Phys. Rev. C {\bf 80}, 054313 (2009).}

\bibitem{[Sch10]}
{N. Schunck, J. Dobaczewski, J. Mor\'e, J. McDonnell, W. Nazarewicz, J. Sarich
  and M.V. Stoitsov, Phys. Rev. C {\bf 81} 024316 (2010).}

\bibitem{[Per08]}
{S. Perez-Martin and L.M. Robledo, Phys. Rev. C {\bf 78}, 014304 (2008).}

\bibitem{[Dob09d]}
{J. Dobaczewski, W. Satu{\l}a, B.G. Carlsson, J. Engel, P. Olbratowski, P.
  Powa{\l}owski, M. Sadziak, J. Sarich, N. Schunck, A. Staszczak, M.V.
  Stoitsov, M. Zalewski, and H. Zdu\'nczuk, Comput. Phys. Commun. {\bf 180},
  2361 (2009).}

\bibitem{[Dob95]}
{J. Dobaczewski and J. Dudek, Phys. Rev. C {\bf 52}, 1827 (1995).}

\bibitem{[Sto10]}
{M.V.~Stoitsov, M.Kortelainen, S.K. Bogner, T. Duguet, R.J. Furnstahl, B.
  Gebremariam, and N. Schunck, Phys. Rev. C {\bf 82}, 054307 (2010).}

\bibitem{[Kor10]}
{M. Kortelainen, T. Lesinski, J. Mor\'e, W. Nazarewicz, J. Sarich, N. Schunck,
  M.V. Stoitsov, and S. Wild, Phys. Rev. C {\bf 82}, 024313 (2010).}

\bibitem{[Kor12]}
{M. Kortelainen, J. Mcdonnell, W. Nazarewicz, P.-G. Reinhard, J. Sarich, N.
  Schunck, M.V. Stoitsov, and S. Wild, Phys. Rev. C {\bf 85}, 024304 (2012).}

\bibitem{[Dob04a]}
{J. Dobaczewski, M.V. Stoitsov, and W. Nazarewicz, AIP Conference Proceedings
  {\bf 726}, 52 (2004).}

\bibitem{[Pei08a]}
{J.C. Pei, M.V. Stoitsov, G.I. Fann, W. Nazarewicz, N. Schunck, and F.R. Xu,
  Phys. Rev. C {\bf 78}, 064306 (2008).}

\bibitem{[Sch09]}
{N. Schunck, M.V. Stoitsov, W. Nazarewicz, and N. Nikolov, AIP Conf. Proc. {\bf
  1128}, 40 (2009) .}

\bibitem{[Cha98]}
{E. Chabanat, P. Bonche, P. Haensel, J. Meyer, and R. Schaeffer, Nucl. Phys. A
  {\bf 635}, 231 (1998)}.

\bibitem{[Gir83]}
{M. Girod and B. Grammaticos, Phys. Rev. C {\bf 27}, 2317 (1983).}

\bibitem{[Sto98]}
{M.V. Stoitsov, W. Nazarewicz, and S. Pittel, Phys. Rev. C {\bf 58}, 2092
  (1998).}

\bibitem{[Sto03]}
{M.V. Stoitsov, J. Dobaczewski, W. Nazarewicz, S. Pittel, and D. J. Dean, Phys.
  Rev. C {\bf 68}, 054312 (2003).}

\bibitem{[Sto08]}
{M. Stoitsov, N. Michel, and K. Matsuyanagi, Phys. Rev. C {\bf 77}, 054301
  (2008).}

\bibitem{[Mic08]}
{N. Michel, K. Matsuyanagi, and M. Stoitsov, Phys. Rev. C {\bf 78}, 044319
  (2008).}

\bibitem{[Dob97c]}
{J. Dobaczewski and J. Dudek, Comput. Phys. Commun. {\bf 102}, 166 (1997).}

\bibitem{[Vau73]}
{D. Vautherin, Phys. Rev. C {\bf 7}, 296 (1973).}

\bibitem{[Fri85]}
{R. Friesner, Chem. Phys. Lett. {\bf 116}, 39 (1985).}

\end{thebibliography}

\end{document}